\documentclass[12pt]{article}%
\usepackage{amssymb}
\usepackage{amsfonts}
\usepackage{float}
\usepackage{graphicx, epsfig}
\usepackage{amsmath}
\usepackage{amsthm}
\usepackage{morefloats}
\usepackage{placeins}
\usepackage{rotating}
\usepackage{natbib}
\usepackage{url}
\usepackage{afterpage}
\usepackage{booktabs}
\usepackage{amssymb}
\usepackage{amsfonts}
\usepackage{graphicx, epsfig}
\usepackage{amsmath}
\usepackage{graphicx}
\usepackage{color}
\usepackage{caption}
\usepackage{url}
\usepackage[utf8]{inputenc}
\usepackage{enumerate}%
\providecommand{\U}[1]{\protect\rule{.1in}{.1in}}

\newtheorem{theorem}{Theorem}

\newtheorem{lemma}{Lemma}

\begin{document}

\begin{center}
{\Large Robust Model-Based Clustering} \medskip

{\large Juan D. Gonz\'{a}lez$^{1}$ ,Ricardo Maronna$^{2}$, Victor J.
Yohai$^{3}$ and Ruben H. Zamar$^{4}$} \medskip

{\large $^{1}$Acoustic Propagation Department-UNIDEF/CONICET, $^{2}$Universidad de Buenos Aires, $^{3}%
$Universidad de Buenos Aires and CONICET and $^{4}$University of British
Columbia}
\end{center}

\section*{Abstract}

{\normalsize We propose a class of Fisher-consistent robust estimators for
mixture models. These estimators are then used to build a \emph{robust
model-based clustering} procedure. We study in detail the case of multivariate
Gaussian mixtures and
propose an algorithm, similar to the EM algorithm, to compute the proposed
estimators and build the robust clusters. An extensive Monte Carlo simulation
study shows that our proposal outperforms other robust and non robust, state
of the art, model-based clustering procedures. We apply our proposal to a real
data set and show that again it outperforms alternative procedures.}

\textit{Keywords:} mixture models, EM--algorithm, scatter S estimators

\section{{\protect\normalsize Introduction}}

{\normalsize Let $f(\mathbf{x},\boldsymbol{\theta)}$} {
be a $p$-dimensional density function indexed by a $q$-dimensional parameter
$\boldsymbol{\theta}$, and let $F_{\boldsymbol{\theta}}(\mathbf{x})$ be the
corresponding distribution function. We consider the mixture model
\begin{equation}
h(\mathbf{x,}\boldsymbol{\alpha},{\boldsymbol{\Theta}})=\sum_{k=1}^{K}%
\alpha_{k}f(\mathbf{x},\boldsymbol{\theta}_{k}), \label{mixture}%
\end{equation}
where $K>0$ is some given integer, $\boldsymbol{\alpha}$$=\left(  {\alpha}%
_{1},...,{\alpha}_{K}\right)  \in\mathbb{[}0,1\mathbb{]}^{K}$, $\sum_{k=1}%
^{K}\alpha_{k}=1$, and $\boldsymbol{\Theta}=\left(  \boldsymbol{\theta}%
_{1},...,\boldsymbol{\theta}_{K}\right)  \in\mathbb{R}^{q\times K}$. We assume
that we have $n$ independent observations from model (\ref{mixture}). The
important case of Gaussian mixtures is obtained when the \textit{kernel
density} $f(\mathbf{x},\boldsymbol{\theta)}$  is a multivariate normal density
with mean $\boldsymbol{\mu}$ and covariance matrix $\boldsymbol{\Sigma}$ (that
is $\boldsymbol{\theta}\boldsymbol{=}\left(  \boldsymbol{\mu}%
\boldsymbol{,\Sigma}\right) $). }

{\normalsize
The seminal work by \cite{dempster1977maximum} introduced the EM algorithm to
compute the maximum likelihood estimators (MLE) for the parameters of a
Gaussian mixture with $K$ components. The MLE is efficient when applied to
clean Gaussian data but performs poorly in the presence of \textit{cluster
outliers}, that is, data points that are far away from all the clusters (see
\cite{garcia2010review}). Several authors addressed the problem of robust
estimation of the parameters of a Gaussian mixture. \cite{garcia2008general}
proposed the maximization of the likelihood of a multivariate normal mixture
after trimming a given fraction, }$\alpha${\normalsize , of the data. This
procedure has a very good performance when the fraction }$\alpha$
{\normalsize is well specified. However, $\alpha$ is often unknown and
difficult to estimate. Another approach, that builds on previous work by
\cite{banfieldRaftery},
was proposed by \cite{corettoHennig2016} and \cite{CHjmlr}. This approach
consists on the addition of an improper uniform distribution with level
$\delta$ to account for possible outliers. In the first implementation of this
procedure called RIMLE the level $\delta$ is a fixed input parameter. In the
current implementation, called OTRIMLE, $\delta$ is estimated from the data.}

{\normalsize We present a new estimating procedure with some desirable
properties: (i) the estimators of the mixture model parameters are
Fisher-consistent and (ii) our method doesn't require prior knowledge of the
fraction of outliers in the data. }

{\normalsize The rest of the paper is organized as follows. In Section
\ref{methodology} we present a general framework for the robust estimation fof
the parameters of a mixture model. In Section \ref{normal mixtures} this
general framework is applied to the case of multivariate Gaussian mixtures. In
Section \ref{algorithm normal} we give a computing algorithm. In Section
\ref{Allocation to clusters} we discuss several practical issues including the
allocation of observations to clusters and the flagging of outliers. In
Section \ref{simulation Study} we present the results of a simulation study.
In Section \ref{real} we apply our clustering procedure to a real dataset and
compare the results with those of alternative cluster procedures. In Section
\ref{conlusions} we give some concluding remarks. Mathematical proofs and
further details are given in the Appendix. }

\section{{\protect\normalsize A General Framework for Robust Estimation of
Mixture Models}}

{\normalsize \label{methodology} }

{\normalsize
}

{\normalsize We consider the problem of robust estimation of the parameters of
the mixture model (\ref{mixture}), $\left(  \boldsymbol{\alpha}%
,\boldsymbol{\Theta}\right)  $, using a random sample $\mathbf{x}
_{1},....\mathbf{x}_{n}$ from this model. }

{\normalsize First we give some general background and context for our
proposal. We can think of model (\ref{mixture}) as the marginal density of an
observation, $\mathbf{X}$, from a random experiment with outcome $\left(
\mathbf{U,X}\right)  $, where the conditional density of $\mathbf{X}$ given
$\mathbf{U=u}$ is $p\left(  \mathbf{x,{\boldsymbol{\Theta}}|U=u}\right)
={\prod_{j=1}^{K}}\left[  f(\mathbf{x},\boldsymbol{\theta}_{j})\right]
^{u_{j}}$ and the label vector $\mathbf{U}$ has multinomial distribution
Mult(}${\normalsize K}${\normalsize ,$\boldsymbol{\alpha}$). Therefore, the
joint density of $\left(  \mathbf{U,X}\right)  $ is }$p\left(  \mathbf{u,x,}%
\boldsymbol{\alpha}\mathbf{,}{\boldsymbol{\Theta}}\right)  ={\prod_{j=1}^{K}%
}\left[  \alpha_{j}f\left(  \mathbf{x},\boldsymbol{\theta}_{j}\right)
\right]  ^{u_{j}}$.


As in the classical EM algorithm, a key building block in the proposed robust
estimation framework is the conditional probability that an observation
$\mathbf{X}$ comes from the $k^{th}$ population given that $\mathbf{X=x}$:
\begin{equation}
\widetilde{\alpha}_{k}(\mathbf{x},\boldsymbol{\alpha,\Theta})=\frac{\alpha
_{k}f(\mathbf{x},\boldsymbol{\theta}_{k})}{\sum_{j=1}^{K}\alpha_{j}f\left(
\mathbf{x},\boldsymbol{\theta}_{j}\right)  }. \label{198}%
\end{equation}
{\normalsize Another key building block is the robust base estimator discussed
below.} Finally, given the robust estimators $(\widehat{\boldsymbol{\alpha}%
},\widehat{\boldsymbol{\Theta}})$ produced by our proposal,
observation $\mathbf{x}_{i}$, $i=1,...,n$, is assigned to cluster $G_{k}$
\ iff $\widetilde{\alpha}_{k}(\mathbf{x}_{i},\widehat{\boldsymbol{\alpha}%
}\boldsymbol{,}\widehat{\boldsymbol{\Theta}})=\max_{1\leq j\leq K}%
\widetilde{\alpha}_{j}(\mathbf{x}_{i},\widehat{\boldsymbol{\alpha}%
}\boldsymbol{,}\widehat{\boldsymbol{\Theta}})$. \newpage

\subsection{{\protect\normalsize The Base Robust Estimator \label{base0} }}

{\normalsize \ }

{\normalsize We assume that given a random sample }$\mathbf{x}_{1}%
,..,\mathbf{x}_{n}$ {\normalsize from the kernel density $f\left(
\mathbf{x},\boldsymbol{\theta}\right)  $, the parameter $\boldsymbol{\theta}$}
\ {\normalsize has a robust estimator $\widehat{\boldsymbol{\theta}}$, which
can be expressed as a function of $h$ sample averages and satisfies a fixed
point equation. More precisely, there exist a function $\mathbf{g}%
:R^{h}\rightarrow R^{q}$ \ and }$h$ real valued functions {\normalsize $\eta
_{h}\left(  \mathbf{x}_{i},\boldsymbol{\theta}\right)  $, \ $1\leq j\leq h$,
such that }%
\begin{equation}
{\normalsize \widehat{\boldsymbol{\theta}}}=\mathbf{g}\left(  \frac{1}{n}%
\sum_{i=1}^{n}\eta_{1}\left(  \mathbf{x}_{i}%
,{\normalsize \widehat{\boldsymbol{\theta}}}\right)  ,...,\frac{1}{n}%
\sum_{i=1}^{n}\eta_{h}\left(  \mathbf{x}_{i}%
,{\normalsize \widehat{\boldsymbol{\theta}}}\right)  \right)  . \label{eq153}%
\end{equation}
In this case, the corresponding asymptotic functional $\boldsymbol{\theta}%
${\normalsize $(F)$ }for {\normalsize $\widehat{\boldsymbol{\theta}}$ \ when
the underlying distribution is }$F$ {\normalsize satisfies the fixed point
equation }%
\begin{equation}
\boldsymbol{\theta}{\normalsize (F)}\boldsymbol{\mathbf{=}}\mathbf{g}\left(
E_{F}\left\{  \eta_{1}\left(  \mathbf{x},\boldsymbol{\theta}{\normalsize (F)}%
\right)  \right\}  ,...,E_{F}\left\{  \eta_{h}\left(  \mathbf{x}%
,\boldsymbol{\theta}{\normalsize (F)}\right)  \right\}  \right)  .
\label{eq154}%
\end{equation}


Many robust estimators satisfy this requirement. \ 

\textbf{Example:} For simplicity's sake, let us consider a univariate location
M-estimator $\widehat{\theta}_{n}$ implicitly defined by the estimating
equation%
\[
\frac{1}{n}\sum_{i=1}^{n}\psi(x_{i}-\widehat{\theta})=0.
\]
To express $\widehat{\theta}$ as in (\ref{eq153}) we write
\[
\frac{1}{n}\sum_{i=1}^{n}\frac{\psi(x_{i}-\widehat{\theta})}{x_{i}%
-\widehat{\theta}}\left(  x_{i}-\widehat{\theta}\right)  =0.
\]
Setting $W\left(  x\right)  =\psi(x)/x$ \ (defined by $\lim_{x\rightarrow
0}\psi(x)/x$ when $x=0$) we have
\[
\frac{1}{n}\sum_{i=1}^{n}W(x_{i}-\widehat{\theta})\left(  x_{i}%
-\widehat{\theta}\right)  =0
\]
or equivalently
\[
\widehat{\theta}_{n}=\frac{\sum_{i=1}^{n}W(x_{i}-\widehat{\theta})x_{i}}%
{\sum_{i=1}^{n}W(x_{i}-\widehat{\theta})}.
\]
This satisfies (\ref{eq153}) with $\eta_{1}\left(  x,\theta\right)
=W(x-\theta)x$, $\eta_{2}\left(  x,\theta\right)  =W\left(  x,\theta\right)  $
and $g(u,v)=u/v$. {\normalsize Similarly, the (more realistic) case of
simultaneous location and scale M-estimators (see Huber, 1964) can also be
written as (\ref{eq153})}. In fact, {\normalsize many robust estimators
satisfy (\ref{eq153}) and (\ref{eq154}).
In particular, we show in Section \ref{normal mixtures} that
\cite{davies1987asymptotic} S estimators of multivariate location and scatter
satisfy these conditions and therefore can be used for the robust estimation
of the parameters of a multivariate Gaussian mixture. }
%

\subsection{{\protect\normalsize The Mixture Model Estimator \label{CEF}}}

Suppose now that {\normalsize we have a robust base estimator
$\widehat{\boldsymbol{\theta}}$ satisfying (\ref{eq153}) and (\ref{eq154}).
Then, given a random sample $\mathbf{x}_{1},..,\mathbf{x}_{n},$ from model
(\ref{mixture}) we define the estimators
\[
\left(  \widehat{\boldsymbol{\alpha}}\mathbf{\boldsymbol{,}%
\widehat{\boldsymbol{\Theta}}}\right)  ,\text{ }\widehat{\boldsymbol{\alpha}%
}\mathcal{=(}\widehat{\mathcal{\alpha}}_{1},...,\widehat{\alpha}_{K}),\text{
}\ \widehat{\mathbf{\boldsymbol{\Theta}}}\boldsymbol{=(}%
\widehat{\boldsymbol{\theta}}_{1},...,\widehat{\boldsymbol{\theta}}_{K})
\]
for the mixture model parameters $\boldsymbol{\alpha}\mathcal{=(\alpha}%
_{1},...,\alpha_{K})$ and \textbf{$\boldsymbol{\Theta}$}$\boldsymbol{=(}%
$$\boldsymbol{\theta}$$_{1},\dots,\boldsymbol{\theta}_{K})$ as follows. Given
$\mathbf{X=x}$, let $\widetilde{\alpha}_{k}(\mathbf{x},\boldsymbol{\alpha
,\Theta})$ be the conditional probability} {\normalsize that this observation
comes from the $k^{th}$sub-population, as given in equation (\ref{198}). Then
$\ $ $\widehat{\alpha}_{k}$ and $\widehat{\boldsymbol{\theta}}_{k\text{ }}%
$\ satisfy the fixed point equations:
\begin{equation}
\widehat{\alpha}_{k}=\frac{1}{n}\sum_{i=1}^{n}\widetilde{\alpha}%
_{k}(\mathbf{x}_{i},\widehat{\boldsymbol{\alpha}}\boldsymbol{,}%
\widehat{\boldsymbol{\Theta}}),\text{ }1\leq k\leq K, \label{eq20a}%
\end{equation}%
\begin{equation}
\widehat{\boldsymbol{\theta}}_{k}=g\left(  \sum_{i=1}^{n}\frac
{\widetilde{\alpha}_{k}(\mathbf{x}_{i},\widehat{\boldsymbol{\alpha}%
}\boldsymbol{,}\widehat{\boldsymbol{\Theta}})}{\widehat{\alpha}_{k}}\eta
_{1}(\mathbf{x}_{i},\widehat{\boldsymbol{\theta}}_{k}),...,\sum_{i=1}^{n}%
\frac{\widetilde{\alpha}_{k}(\mathbf{x}_{i},\widehat{\boldsymbol{\alpha}%
}\boldsymbol{,}\widehat{\boldsymbol{\Theta}})}{\widehat{\alpha}_{k}}\eta
_{h}(\mathbf{x}_{i},\widehat{\boldsymbol{\theta}}_{k})\right)  ,\text{ }1\leq
k\leq K, \label{eq20}%
\end{equation}
respectively. }

Notice that $\widehat{\boldsymbol{\theta}}_{k}$ is the base estimator defined
in (\ref{eq153}) (still using the $n$ observations) but with simple averages
replaced by weighted averages. The $i^{th}$ observation $\mathbf{x}_{i}$ has a
weight proportional to the conditional probability, $\widetilde{\alpha}%
_{k}(\mathbf{x}_{i},\boldsymbol{\alpha,\Theta})$, that $\mathbf{x}_{i}$ belong
to the $k^{th}$ sub-population.

{\normalsize Given the mixed model distribution $H$ we denote by
$\mathbf{T(}H\mathbf{)=}\left(  \boldsymbol{\alpha}\left(  H\right)
\boldsymbol{,\Theta}\left(  H\right)  \right)  $ the corresponding asymptotic
functional} of the robust estimators. \ The {\normalsize $K$\ components
\ of\ $\boldsymbol{\alpha}\left(  H\right)  $\textbf{ }and\textbf{
}$\boldsymbol{\Theta}\left(  H\right)  $ satisfy the fixed point equations }

{\normalsize
\begin{equation}
\alpha_{k}=\ E_{H}\left(  \widetilde{\alpha}_{k}(\mathbf{x}_{i}%
,\mathcal{\boldsymbol{\alpha}}\boldsymbol{,\Theta})\right)  ,\text{ }1\leq
k\leq K, \label{eqalfa}%
\end{equation}
\bigskip%
\begin{equation}
\boldsymbol{\theta}_{k\text{ }}=g\left(  E_{H}\left(  \frac{\widetilde{\alpha
}_{k}(\mathbf{x},\boldsymbol{\alpha,\Theta})}{\alpha_{k}}\eta_{1}%
(\mathbf{x},\boldsymbol{\theta}_{k})\right)  ,...,E_{H}\left(  \frac
{\widetilde{\alpha}_{k}(\mathbf{x},\boldsymbol{\alpha,\Theta})}{\alpha_{k}%
}\eta_{h}(\mathbf{x},\boldsymbol{\theta}_{k})\right)  \right)  ,\text{ }1\leq
k\leq K \label{eqtheta}%
\end{equation}
} \medskip{\normalsize The theorem below shows that if the robust base
estimator }${\normalsize \widehat{\boldsymbol{\theta}}}$ {\normalsize is
Fisher consistent, that is, if the corresponding asymptotic functional
}$\boldsymbol{\theta}\left(  F\right)  $ satisfies the equation {\normalsize
\begin{equation}
\boldsymbol{\theta\mathbf{=}}\mathbf{g}\left(  E_{F_{\boldsymbol{\theta}}%
}\left\{  \eta_{1}\left(  \mathbf{x},\boldsymbol{\theta}\left(
F_{\boldsymbol{\theta}}\right)  \right)  \right\}
,...,E_{F_{\boldsymbol{\theta}}}\left\{  \eta_{h}\left(  \mathbf{x}%
,\boldsymbol{\theta}(F_{\boldsymbol{\theta}})\right)  \right\}  \right)
,\text{ \ for all }\boldsymbol{\theta,} \label{eq54.+}%
\end{equation}
then the estimators for the mixture distribution parameters proposed above are
also Fisher consistent. }

{\normalsize
}

\begin{theorem}
{\normalsize \label{Th 1} Suppose that }$\boldsymbol{\Theta}$%
{\normalsize $_{0}=($$\boldsymbol{\theta}$$_{01},...,$ $\boldsymbol{\theta}%
$$_{0K})$ and $\boldsymbol{\alpha}$$_{0}=(\alpha_{01},...,\alpha_{0K})$ are
the true values of }$\boldsymbol{\Theta}${\normalsize and $\boldsymbol{\alpha
}$, respectively. Let $H_{0}$ be the corresponding true mixture distribution
with density
\[
h_{0}(\mathbf{x,}\boldsymbol{\alpha}_{0},\boldsymbol{\Theta}_{0}%
)=\sum\limits_{k=1}^{K}\alpha_{0k}f(\mathbf{x,}\boldsymbol{\theta}_{0k}).
\]
Suppose that the base estimator $\widehat{\boldsymbol{\theta}}$ is Fisher
consistent, then }$\left(  \widehat{\boldsymbol{\alpha}}\mathbf{\boldsymbol{,}%
\widehat{\boldsymbol{\Theta}}}\right)  $ {\normalsize is also Fisher
consistent. That is }%
\[
{\normalsize \mathbf{T}(H_{0}\mathcal{)}=(\boldsymbol{\alpha}(H_{0}%
\mathcal{)},\boldsymbol{\Theta}(H}_{0}{\normalsize \mathcal{)})=}\left(
{\normalsize \boldsymbol{\alpha}}_{0}{\normalsize ,\boldsymbol{\Theta}_{0}%
}\right)  {\normalsize ,}\text{ \ \ for all }{\normalsize (\boldsymbol{\alpha
}_{0},\boldsymbol{\Theta}_{0}).}%
\]

\end{theorem}

{\normalsize
}

\subsection{{\protect\normalsize Computing Strategy \ \label{COMPstrat}}}

{\normalsize \ Let $\mathbf{x}_{1},...,\mathbf{x}_{n}$ be a random sample from
the mixture model (1) and let $H_{n}$ be the corresponding empirical
distribution function. We compute estimators $(\widehat{\boldsymbol{\alpha}%
},\widehat{\boldsymbol{\theta}})=\mathbf{T}\mathcal{(}H_{n})$ using an
iterative approach. Suppose that, at step $m$, the current values of the
estimators are $\boldsymbol{\alpha}^{m}=\left(  \alpha_{1}^{m},...,\alpha
_{K}^{m}\right)  $ and $\boldsymbol{\Theta}^{m}=(\boldsymbol{\theta}_{1}%
^{m},...,\boldsymbol{\theta}_{K}^{m})$. Then, for $1\leq k \leq K$, we set }

{\normalsize
\[
\alpha_{k}^{m+1}=E_{H_{n}}(\widetilde{\alpha}_{k}(\mathbf{x}%
,\boldsymbol{\alpha}^{m}\boldsymbol{,\Theta}^{m})),\ 1\leq k\leq K,
\]
and%
\[
\boldsymbol{\theta}_{k}^{m+1}=\boldsymbol{g}\left(  E_{H_{n}}\left(
\frac{\widetilde{\alpha}_{k}(\mathbf{x},\boldsymbol{\alpha}^{m}%
\boldsymbol{,\Theta}^{m})}{\alpha_{k}^{m+1}}\eta_{1}(\mathbf{x}%
,\boldsymbol{\theta}_{k}^{m})\right)  ,...,E_{H_{n}}\left(  \frac
{\widetilde{\alpha}_{k}(\mathbf{x},\boldsymbol{\alpha}^{m}\boldsymbol{,\Theta
}^{m})}{\alpha_{k}^{m+1}}\eta_{h}(\mathbf{x},\boldsymbol{\theta}_{k}%
^{m})\right)  \right)  .
\]
}{\normalsize \ Observe that if $(\boldsymbol{\alpha}^{m}\boldsymbol{,\Theta
}^{m})\rightarrow(\boldsymbol{\alpha,\Theta}),$ then $(\boldsymbol{\alpha
,\Theta})$ satisfies the fixed point equations (\ref{eq20a}) and
\ (\ref{eq20}).\bigskip\ }

{\normalsize \noindent\textbf{Initial estimators } One way to define the
initial estimators $\boldsymbol{\alpha}^{0}$ and $\boldsymbol{\Theta}^{0}$ for
a multivariate normal mixture is given in Section \ref{algorithm normal}. }

{\normalsize \noindent\textbf{Stopping rule.} For each $m$, let $H^{m}$ be the
mixture model distribution with $(\boldsymbol{\alpha},\Theta
)=(\boldsymbol{\alpha}^{m},\Theta^{m})$. We stop the iterations when $H^{m}$
and $H^{m+1}$ are close enough. See Section \ref{algorithm normal} for further
details for the case of multivariate normals mixtures. }

{\normalsize \ }

{\normalsize
}

\section{{\protect\normalsize Robust Estimation of Normal Mixtures
\label{normal mixtures}}}

{\normalsize In this section we propose a robust estimator for the parameters
of a multivariate normal mixture model, based in the estimators defined in
Section \ref{CEF}. In this case the kernel density (\ref{mixture}) is a
multivariate normal with mean $\boldsymbol{\mu}\text{ and covariance matrix
}\Sigma$ and the chosen robust base estimator is the S estimator for
multivariate location and scatter matrix (\cite{davies1987asymptotic}),
defined as follows.
Given a $p$-dimensional vector $\boldsymbol{\mu}$, a $p\times p$ symmetric and
positive definite matrix $\Sigma$, and a distribution $F$ on $\mathbb{R}^{p}$,
the asymptotic scale functional $\sigma\boldsymbol{(}F,\boldsymbol{\mu}%
,\Sigma)$ is implicitly defined by the equation
\[
E_{F}\left(  \rho_{c}\left(  \frac{d(\mathbf{x},\boldsymbol{\mu},\Sigma
)}{\sigma\boldsymbol{(}F,\boldsymbol{\mu},\Sigma)}\right)  \right)  =b,
\]
with
\begin{equation}
d^{2}(\mathbf{x},\boldsymbol{\mu},\Sigma)=(\mathbf{x}-\boldsymbol{\mu
})^{\text{T}}\Sigma^{-1}(\mathbf{x}-\boldsymbol{\mu}), \label{defisig}%
\end{equation}
where $0.5\leq b\leq1$ and $\rho_{c}(d)=\rho(d/c)$, for a non-negative \ and
non-decreasing function $\rho$ such that $\rho(0)$=0 and $\sup\rho(d)=1$. The
tuning constant $c>0$ is chosen so that
\begin{equation}
E(\rho_{c}(Y^{1/2}))=b,\text{ \ \ \ }Y\thicksim\chi_{(p)}^{2}.
\label{formulaS}%
\end{equation}
\ Then if $F$ is $N(\boldsymbol{\mu},\Sigma)$ we have $\sigma\boldsymbol{(}%
F,$$\boldsymbol{\mu}$$,\Sigma)=1$. The value of $b$ determines the breakdown
point of the estimator which is equal to $\min(b,1-b)$. Finally, the S
estimator functional of multivariate location and scatter is defined by
\begin{equation}
(\boldsymbol{\mu}(F),\Sigma(F))=\arg\min_{\sigma\boldsymbol{(}%
F,\boldsymbol{\mu},\Sigma)=1}|\Sigma|, \label{defSestimador}%
\end{equation}
where $|\Sigma|$ denotes the determinant of $\Sigma$. }


Given a sample $\mathbf{x}_{1},...,\mathbf{x}_{n}$ in $\ \mathbb{R}^{p}$, the
S estimator of multivariate location and scatter is obtained replacing $F$ by
the empirical distribution $F_{n}$. That is, 

{\normalsize
\begin{equation}
(\boldsymbol{\mu}(F_{n}),\Sigma(F_{n}))=\arg\min_{\sigma(F_{n},\boldsymbol{\mu
},\Sigma)=1}|\Sigma|, \label{S-estimador}%
\end{equation}
with $\sigma\boldsymbol{(}F_{n},\boldsymbol{\mu,\Sigma)}$ given by the
equation
\[
\frac{1}{n}\sum_{i=1}^{n}\rho_{c}\left(  \frac{d(\mathbf{x}_{i}%
,\boldsymbol{\mu},\Sigma)}{\sigma(F_{n},\boldsymbol{\mu},\Sigma)}\right)  =b.
\]
}

\subsection{{\protect\normalsize S estimators Fit the General Framework of
Section \ref{base0}}}

{\normalsize To write the asymptotic $S$ functional as a fixed point of a
function of means we need to introduce the auxiliary parameters $\Sigma^{\ast
}$ and $s^{\ast}$. The fixed point equation satisfied by the augmented S
functional $\ $$(\boldsymbol{\mu}(F),\Sigma(F),s^{\ast}(F),\Sigma^{\ast}(F))$
is given in the following theorem.  }

\begin{theorem}
{\normalsize \label{fixpointequation} Let $\psi=\rho^{\prime}$ and
$W(t)=\psi(d)/d$. Let $($$\boldsymbol{\mu}$$(F),\Sigma(F))$ be the S
functional, then there exists a $p\times p$ symmetric and positive definite
matrix $\Sigma^{\ast}(F)$ and a scalar $s^{\ast}(F)$\ such that
$(\boldsymbol{\mu}(F),\Sigma(F),s^{\ast}(F),\Sigma^{\ast}(F))$ satisfies the
following fixed point equations
\begin{align*}
\boldsymbol{\mu}(F)  &  =\frac{E_{F}\left(  W\left(  d(\mathbf{x}%
,\boldsymbol{\mu}(F),\Sigma(F))\right)  \mathbf{x}\right)  }{E_{F}\left(
W\left(  d(\mathbf{x},\boldsymbol{\mu}(F),\Sigma(F))\right)  \right)  },\\
\Sigma^{\ast}(F)  &  =\frac{E_{F}\left(  W\left(  d(\mathbf{x},\boldsymbol{\mu
}(F),\Sigma(F)\right)  (\mathbf{x}-\boldsymbol{\mu}(F))(\mathbf{x}%
-\boldsymbol{\mu}(F))^{\text{T}}\ \right)  }{E_{F}\left(  W\left(
d(\mathbf{x},\boldsymbol{\mu}(F),\Sigma(F))\right)  \right)  },\\
s^{\ast}(F)  &  =E_{F}\left(  2s^{\ast}(F)\rho\left(  d(\mathbf{x}%
,\boldsymbol{\mu}\left(  F\right)  ,\Sigma^{\ast}(F)/s^{\ast}(F)\right)
\right)  ,\\
\Sigma(F)  &  =s^{\ast}(F)^{2}\Sigma^{\ast}(F).\\
&
\end{align*}
}
\end{theorem}

{\normalsize Theorem 2 shows that the augmented S functional $(\boldsymbol{\mu
}(F),\Sigma(F),s^{\ast}(F),\Sigma^{\ast}(F))$ satisfies the requirements
specified for the base estimating functional given in Section 2 with
\begin{equation}%
\begin{array}
[c]{ll}%
\eta_{1}(\mathbf{x},\boldsymbol{\mu},\Sigma^{\ast},s^{\ast},\Sigma) &
=W\left(  d(\mathbf{x},\boldsymbol{\mu},\Sigma)\right)  \mathbf{x,}\\
\eta_{2}(\mathbf{x},\boldsymbol{\mu},\Sigma^{\ast},s^{\ast},\Sigma) &
=W\left(  d(\mathbf{x},\boldsymbol{\mu},\Sigma)\right)  ,\\
\eta_{3}(\mathbf{x},\boldsymbol{\mu},\Sigma^{\ast},s^{\ast},\Sigma) &
=W\left(  d(\mathbf{x},\boldsymbol{\mu},\Sigma\right)  )(\mathbf{x}%
-\boldsymbol{\mu})(\mathbf{x}-\boldsymbol{\mu})^{\text{T}},\\
\eta_{4}(\mathbf{x},\boldsymbol{\mu},\Sigma^{\ast},s^{\ast},\Sigma) &
=2s^{\ast}\rho\left(  d\left(  \mathbf{x},\boldsymbol{\mu},\Sigma^{\ast
}\right)  /s^{\ast}\right)  ,
\end{array}
\end{equation}
}

{\normalsize
\[
\mathbf{g}(z_{1},z_{2},z_{3},z_{4})=\left(  z_{1}/z_{2},z_{3}/z_{2}%
,z_{4},z_{4}^{2}z_{3}/z_{2}\right)  .
\]
and fixed point equations  }

{\normalsize
\begin{equation}%
\begin{array}
[c]{ll}%
\boldsymbol{\mu} & =E(\eta_{1}(\mathbf{x},\boldsymbol{\mu},\Sigma^{\ast
},\Sigma,s^{\ast}))/E(\eta_{2}(\mathbf{x},\boldsymbol{\mu},\Sigma^{\ast
},\Sigma,s^{\ast})),\\
\Sigma^{\ast} & =E(\eta_{3}(\mathbf{x},\boldsymbol{\mu},\Sigma^{\ast}%
,\Sigma,s^{\ast}))/E(\eta_{2}(\mathbf{x},\boldsymbol{\mu},\Sigma^{\ast}%
,\Sigma,s^{\ast}),\\
s^{\ast} & =E(\eta_{4}(\mathbf{x},\boldsymbol{\mu},\Sigma^{\ast}%
,\Sigma,s^{\ast})),\\
\Sigma & =E(\eta_{4}(\mathbf{x},\boldsymbol{\mu},\Sigma^{\ast},\Sigma,s^{\ast
}))^{2}E(\eta_{3}(\mathbf{x},\boldsymbol{\mu},\Sigma^{\ast},\Sigma,s^{\ast
}))/E(\eta_{2}(\mathbf{x},\boldsymbol{\mu},\Sigma^{\ast},\Sigma,s^{\ast}).
\end{array}
\label{fp}%
\end{equation}
}

\subsection{{\protect\normalsize The Loss Function}}

{\normalsize In this paper we use the loss function
\begin{equation}
\rho(t)=\left\{
\begin{array}
[c]{lcc}%
1.38t^{2} & \text{if}\  & 0\leq t<2/3\\
0.55-2.69t^{2}+10.76t^{4}-11.66t^{6}+4.04t^{8} & \text{if} & 2/3\leq|t|\leq1\\
1 & \text{if} & |t|>1.
\end{array}
\right.  \label{rhoc}%
\end{equation}
This is a simplified version of the optimal $\rho$ function obtained by
\cite{yohai1997optimal} for robust regression. Simulation studies showed that
the S estimators for multivariate location and scatter based on these type of
$\rho$ functions have better performance than those based on the more
traditional Tukey bisquare loss function (see \cite{maronna2017robust}). For
the remainder of this work we take $b=0.5$ which is the value maximizing the
breakdown point.
To simplify the notation, in the following we write $\rho$ instead of
$\rho_{c}$.  }

{\normalsize The values of $c$ that satisfy equation (\ref{formulaS}) with
$b=0.5$ for $\rho$ functions given in (\ref{rhoc}) can be found in Table
\ref{tablavaloresc} for $1\leq p\leq$ $20$. An approximation (good for $p$ in
the range $1\leq p\leq400$) is given by
\[
\hat{c}(p)=-\frac{0.1642}{p}+0.5546\sqrt{p}.
\]
The maximum error of this approximation is 0.015. That is, $|\hat
{c}(p)-c(p)|\leq0.015$ for $1\leq p\leq400$.  }

{\normalsize \begin{table}[ptb]
\begin{center}
{\normalsize \
\begin{tabular}
[c]{ccccccccccc}\hline
$p$ & 1 & 2 & 3 & 4 & 5 & 6 & 7 & 8 & 9 & 10\\\hline
$c $ & 1.21 & 2.08 & 2.70 & 3.19 & 3.61 & 3.99 & 4.33 & 4.65 & 4.94 &
5.22\\\hline
\end{tabular}
}
\par
{\normalsize \bigskip}
\par
{\normalsize \
\begin{tabular}
[c]{ccccccccccc}\hline
$p$ & 11 & 12 & 13 & 14 & 15 & 16 & 17 & 18 & 19 & 20\\\hline
$c$ & 5.48 & 5.73 & 5.97 & 6.20 & 6.42 & 6.64 & 6.84 & 7.04 & 7.24 &
7.43\\\hline
\end{tabular}
}
\end{center}
\par
{\normalsize \  }\caption{Value of the tuning constants satisfying equation
(\ref{formulaS}) for different values of $p$.}%
\label{tablavaloresc}%
\end{table}\bigskip}

\section{{\protect\normalsize Computing Algorithm \label{algorithm normal}}}

{\normalsize We now apply the computing strategy described in Section
\ref{COMPstrat} to the case of mixture of Gaussian distributions with fixed
point equations \ (\ref{fp}).  }

\begin{description}
\item[Initialization.] {\normalsize We will assume that the number of clusters
$K$ is given. The initial values $\boldsymbol{\mu}^{0}=\left(  \boldsymbol{\mu
}_{1}^{0},...,\boldsymbol{\mu}_{K}^{0}\right)  ,\boldsymbol{\Sigma}%
^{0}=\left(  \boldsymbol{\Sigma}_{1}^{0},...,\boldsymbol{\Sigma}_{K}%
^{0}\right)  $, \ $\boldsymbol{\alpha}$$^{0}=\left(  \alpha_{1}^{0}%
,...,\alpha_{K}^{0}\right)  $ and $s^{\ast0}$ can be obtained as
follows:\linebreak}
\end{description}

{\normalsize \noindent\textbf{Initial estimator for} $\boldsymbol{\mu}_{k}$:
we use the K-Tau estimator for the cluster centers given by
\cite{paperArxivKTAU}. \smallskip}

{\normalsize \noindent\textbf{Initial estimator for} $\boldsymbol{\alpha}$: we
first make an initial assignment of the data points to sub-populations by
minimizing their Euclidean distances to the initial cluster centers
$\boldsymbol{\mu}_{k}$. The initial values for the $\alpha_{k}$ are then taken
equal to the relative frequency of each sub-population. \smallskip}

{\normalsize \noindent\textbf{Initial estimator for} ${\Sigma}_{k}$: we use
the points assigned to each sub-population to compute the robust estimator of
scatter proposed by \cite{davies1987asymptotic}. \smallskip}

\begin{description}
\item[Iteration.] {\normalsize
Let $\boldsymbol{\alpha}^{m}$, $\boldsymbol{\mu}^{m}$ and $\boldsymbol{\Sigma
}^{m}$, be the current values for the mixture parameters then
$\boldsymbol{\alpha}^{m+1}$ and $\boldsymbol{\mu}^{m+1},\Sigma^{m+1}$ are
computed as follows.  }

\begin{description}
\item[(a)] {\normalsize Obtain $\widetilde{\alpha}$$_{ki}$, $1\leq i\leq
n,1\leq k\leq K$, the probability that $\mathbf{x}_{i}$ belong belongs to the
$k^{th}$ sub-population when the mixture model parameters are
$\boldsymbol{\alpha}^{m}$, $\boldsymbol{\mu}^{m}$ and $\boldsymbol{\Sigma}%
^{m}$ \
\begin{equation}
\ \widetilde{\alpha}_{ki}=\frac{f\left(  \mathbf{x}_{i},\boldsymbol{\mu}%
_{k}^{m},\Sigma_{k}^{m}\right)  \alpha_{k}^{m}}{\sum_{l=1}^{K}f\left(
\mathbf{x}_{i},\boldsymbol{\mu}_{l}^{m},\Sigma_{l}^{m}\right)  \alpha_{l}^{m}%
}, \label{eq4est}%
\end{equation}
$\ $where
\[
f\left(  \mathbf{x},\boldsymbol{\mu},\Sigma\right)  =(2\pi)^{-p/2}%
|\Sigma|^{-1/2}e^{-\frac{1}{2}(\mathbf{x}-\boldsymbol{\mu})^{\text{T}}%
\Sigma^{-1}(\mathbf{x}-\boldsymbol{\mu}).}%
\]
}

\item[(b)] {\normalsize Update $\alpha_{k},1\leq k\leq K,$
\begin{equation}
\alpha_{k}^{m+1}=\frac{\sum_{i=1}^{n}\overset{}{\widetilde{\alpha}}_{ki}}{n}.
\end{equation}
}

\item[(c)] {\normalsize Update $\boldsymbol{\mu}_{k},$ $1\leq k\leq K.$ First
we compute $\widetilde{d}$$_{ik}^{{}}=$ $d(\mathbf{x}_{i},$$\boldsymbol{\mu}%
$$_{k}^{m},\Sigma_{k}^{m}),$ $1\leq i\leq n,$ $1\leq k\leq K$ and then
$\boldsymbol{\mu}_{k}^{m+1}$ is the expectation of $\mathbf{x}$ when
$\mathbf{x}_{i},$ $1\leq i\leq n$ has probability $\widetilde{\alpha}$%
$_{ki}W(\widetilde{d}_{ik})/\sum_{i=1}^{n}\widetilde{\alpha}_{ki}%
\ W(\widetilde{d}_{ik}^{{}}),$ then \
\[
\boldsymbol{\mu}_{k}^{m+1}=\frac{\sum_{i=1}^{n}\widetilde{\alpha}%
_{ki}\ W(\widetilde{d}_{ik})\mathbf{x}_{i}\ }{\sum_{i=1}^{n}\widetilde{\alpha
}_{ki}\ W(\widetilde{d}_{ik})}.
\]
}

\item[(d)] {\normalsize Update $\Sigma_{k}^{\ast}$, $1\leq k\leq K$.  $\left(
\Sigma_{k}^{\ast}\right)  ^{m+1}$ is the expectation of $(\mathbf{x}%
-\boldsymbol{\mu}^{m+1})(\mathbf{x}-\boldsymbol{\mu}^{m+1})^{\text{T}}$ when
\ $\mathbf{x}_{i},$ $1\leq i\leq n$, has probability $\widetilde{\alpha}%
$$_{ki}W(\widetilde{d}_{ik})/\sum_{i=1}^{n}\widetilde{\alpha}_{ki}%
\ W(\widetilde{d}_{ik}),$ then \
\[
\left(  \Sigma_{k}^{\ast}\right)  ^{m+1}=\frac{\sum_{i=1}^{n}\widetilde{\alpha
}_{ki}^{{}}W(\widetilde{d}_{ik})(\mathbf{x}_{i}-\boldsymbol{\mu}%
^{m+1})(\mathbf{x}_{i}-\boldsymbol{\mu}^{m+1})^{\text{T}}}{\sum_{i=1}%
^{n}\widetilde{\alpha}_{ki}\ W(\widetilde{d}_{ik})}%
\]
$\ $ }

\item[(e)] {\normalsize Update $s_{k}^{\ast},1\leq k\leq K.$ First we
recompute $\widetilde{d}$$_{ik}=$ $d(\mathbf{x}_{i},$$\boldsymbol{\mu}$%
$_{k}^{m+1},\left(  \Sigma_{k}^{\ast}\right)  ^{m+1}),$ $1\leq i\leq n,$
$1\leq k\leq K.$ Then
\[
\left(  s_{k}^{\ast}\right)  ^{m+1}=2s_{k}^{\ast m}\frac{1}{n}\sum_{i=1}%
^{n}\frac{\widetilde{\alpha}_{ki}}{\alpha_{k}^{m+1}}\rho\left(  \widetilde{{d}%
}{_{ik}}/{s_{k}^{\ast m}}\right)  .
\]
}

\item[(f)] {\normalsize Update $\Sigma_{k},1\leq k\leq K$
\[
\Sigma_{k}^{m+1}=\left[  \left(  s_{k}^{\ast}\right)  ^{m+1}\right]
^{2}\left(  \Sigma_{k}^{\ast}\right)  ^{m+1}.
\]
}
\end{description}

\item[Stopping Rule.] {\normalsize The iterations stop when
\[
\left\Vert {\boldsymbol{\alpha}^{m+1}-\boldsymbol{\alpha}^{m}}\right\Vert
<\delta
\]
and
\[
\sum_{k=1}^{K}d_{KL}(F_{k}^{m+1},F_{k}^{m})<\delta,
\]
where $\delta>0$ is the desired precision and $d_{KL}(F_{k}^{m+1},F_{k}^{m})$
are the Kullback--Leibler divergences between the distributions of the
$k^{th}$ components obtained at iterations $m$ and $m+1$, respectively.  }
\end{description}

{\normalsize \textbf{Remark: } It may happen in some occasions that the matrix
inversion required to compute the distance $d$ defined in (\ref{defisig}) is
not possible because of collinearity. In this case, we notice that these
distances are only needed to compute the probabilities $\widetilde{\alpha
}_{ki}$ in (\ref{eq4est}) and the weights $W(\widetilde{d}_{ik})$ in the
ensuing steps. Fortunately this problem can be resolved using the general
procedure to compute Mahalanobis distances described in Section 6.3.1 of
\cite{Robook19}. }

{\normalsize We now make a conceptual comparison between our algorithm and the
EM algorithm for the case of multivariate normal mixtures. The update of the
mixture weights $\boldsymbol{\alpha}$, steps (a) and (b) of the iteration, are
exactly the same in both algorithms. The updates for $\boldsymbol{\mu}_{k}$,
step (c), are quite similar in both algorithms. In both cases the updating
formulas are weighted means of the observations $\mathbf{x}_{i}$.
However, while the weights used in the EM algorithm are proportional to
$\widetilde{\alpha}$$_{ki}$, the probability that $\mathbf{x}_{i}$ belongs to
the $k^{th}$ sub-population, the weights used in our robust algorithm are
proportional to the products $\widetilde{\alpha}$$_{ki}W(\widetilde{d}_{ik})$.
The extra factor $W(\widetilde{d}_{ik})$ decreases with the distance of
$\mathbf{x}_{i}$ to the center $\boldsymbol{\mu}_{k}$ of the $k^{th}$ mixture
component, ensuring that outliers that are far away from all the cluster
centers have small - even zero - weight and therefore little influence on the
value of the updated estimators of $\boldsymbol{\mu}_{k}$. A similar comment
applies to the update the matrix $\Sigma_{k}^{\ast}$, step (d) in both
algorithms.
Our robust algorithm has two extra steps, steps (e) and (f), which are needed
for a technical reason related to the use of S estimators: the matrix
$\Sigma_{k}^{\ast}$ is slightly biased as an estimator of $\Sigma_{k}$ and
requires a scalar correction factor $(s_{k}^{\ast})^{2}$, which is calculated
in step (e) and used in step (f). These steps are not needed in the case of
the EM algorithm.  }

{\normalsize Notice that if $\rho(d)=ad^{2}$ for some constant $a>0$, then
$W(\widetilde{d}_{ik})=2a$ for all $d_{ik}$, and our algorithm reduces to the
EM algorithm. Moreover, in the case that $\rho(d_{ik}/s^{\ast})=\rho
_{c}(d_{ik})=\rho(d_{ik}/c)$, with $\rho$ given by (\ref{rhoc}), if $c$ is
sufficiently large (as is our recommended default) and there are no outliers,
then $d_{ik}\leq(2/3)c$ for all $i$ and $\rho_{c}(d_{ik})=(1.38/c)d_{ik}^{2}$.
Therefore, when the data  have no outliers the estimators produced by the
robust algorithm and the classical EM algorithm are very similar. However,
when there are outliers, these outliers may gravely affect the EM-algorithm
but not much the robust algorithm because they will be assigned small or even
zero weights.
}

\section{{\protect\normalsize \ Robust Clustering
\label{Allocation to clusters}}}

{\normalsize We can use the robust estimators $\widehat{\boldsymbol{\mu}%
}=\left(  \hat{\boldsymbol{\mu}}_{1},...,\hat{\boldsymbol{\mu}}_{K}\right)
,\hat{\boldsymbol{\Sigma}}=\left(  \hat{{\Sigma}}_{1},...,\hat{{\Sigma}}%
_{K}\right)  $ and $\hat{\mathbf{\boldsymbol{\alpha}}}=\left(  \hat{\alpha
}_{1},...,\hat{\alpha}_{K}\right)  $ to define robust clusters. This approach
is called \emph{robust model-based clustering} (RMBC) . }

{\normalsize \noindent Let
\begin{equation}
\widehat{{P}}(\mathbf{x}_{i}\in G_{k})=\frac{f\left(  \mathbf{x}_{i}%
,\hat{\boldsymbol{\mu}}_{k},\hat{\Sigma}_{k}\right)  \hat{\alpha}_{k}}%
{\sum_{l=1}^{K}f\left(  \mathbf{x}_{i},\hat{\boldsymbol{\mu}}_{l},\hat{\Sigma
}_{l}\right)  \hat{\alpha}_{l}}. \label{perte}%
\end{equation}
}

{\normalsize \noindent Observation $\mathbf{x}_{i}$ is assigned to cluster
$G_{j}$ if }

{\normalsize
\[
\widehat{{P}}(\mathbf{x}_{i}\in G_{j}) > \widehat{{P}}(\mathbf{x}_{i}\in
G_{k}), \ \ \mbox{for all} \ k \neq j.
\]
}

{\normalsize \noindent Since the denominator in (\ref{perte}) is constant over
$k$ we just need to compare the numerators in the log scale:
\[
\delta_{k}(\mathbf{x)=}\log\hat{\alpha}_{k}-\frac{1}{2}\log|\hat{\Sigma}%
_{k}|-\frac{1}{2}d^{2}(\mathbf{x},\hat{\boldsymbol{\mu}}_{k},\hat{\Sigma}%
_{k}).
\]
\medskip
}

{\normalsize \noindent\textbf{Flagging outliers:} Let $\{\mathbf{x}%
\in\mathbb{R}^{p}:d^{2}(\mathbf{x},\hat{\boldsymbol{\mu}}_{k},\hat{\Sigma}%
_{k})\leq\chi_{p,1-\beta}^{2}\}, \label{ellipse2}$
with $\beta=10^{-3}$, and set $\mathcal{E}^{K}=\cup_{k=1}^{K}\mathcal{E}_{k}$.
Observation $\mathbf{x}_{i}$ is flagged as outlier if it falls outside
$\mathcal{E}^{K}$. }

{\normalsize
}

\section{{\protect\normalsize Simulation Study \label{simulation Study}}}

\subsection{{\protect\normalsize Scenarios Used in Our Simulation}}

{\normalsize We generate 500 replications from six different scenarios. In the
first four scenarios the data have a contaminated mixture distribution with
$K$ components. The density has the following form
\begin{equation}
(1-\varepsilon)\sum_{j=1}^{K}\alpha_{k}f(\mathbf{x},\boldsymbol{\mu}%
_{k},\Sigma_{k})+\varepsilon f_{c}(x),.\label{ecuacionMezcla}%
\end{equation}
where $\varepsilon$ is the fraction of contamination and $f_{\varepsilon}$ is
the contamination density The first three scenarios, taken from
\cite{corettoHennig2016}, have fixed covariance matrices and are named
SunSpot5, SideNoise2 and SideNoise2H (as in the given reference). We also
simulated another scenario, SideNoise3, with fixed covariance matrix and two
scenarios called RandScatterMatrix and RandScatterMatrixH, where the
covariance matrices are generated at random for each replication. In the
random covariance scenarios the outliers are generated in a different
appropriate way for each replication. }

{\normalsize \bigskip\noindent{\textbf{SunSpot5}:} In this case we have $K=5$
clusters, with weights $\boldsymbol{\alpha}=(0.15,0.30,0.10,0.15,0.30), $ in
$R^{2}$. The kernel distribution is normal,
\[
\boldsymbol{\mu}_{1}=(0,3)\quad\boldsymbol{\mu}_{2}=(7,1)\quad\boldsymbol{\mu
}_{3}=(5,9)\quad\boldsymbol{\mu}_{4}=(-13,5)\quad\boldsymbol{\mu}_{5}=(-9,5),
\]
\medskip%
\[
\Sigma_{1}=\left(
\begin{array}
[c]{cc}%
1 & 0.5\\
0.5 & 1
\end{array}
\right)  \quad\Sigma_{2}=\left(
\begin{array}
[c]{cc}%
2 & -1.5\\
-1.5 & 2
\end{array}
\right)  \quad\Sigma_{3}=\left(
\begin{array}
[c]{cc}%
2 & 1.3\\
1.3 & 2
\end{array}
\right)  ,
\]
$\Sigma_{4}=0.5I_{2}$ $\ $and $\Sigma_{5}=2.5I_{2}$. In general, $I_{p}$
denotes the identity matrix of dimension $p$. The fraction of contamination is
$\varepsilon=0.005$ with uniform distribution in the rectangle $[30,40]\times
\lbrack30,40]$. The sample size in this case is $n=1000$. This scenario
generates a few isolated outliers far away from the bulk of data. }

{\normalsize \noindent{\textbf{SideNoise2}:} \label{low} \ In this case we
consider $K=2$ clusters with weights $\boldsymbol{\alpha}=(0.75,0.25)$ \ in
$R^{2}\ $. The kernel distribution is normal,
\[
\boldsymbol{\mu}_{1}=(-10,5),\text{ }\boldsymbol{\mu}_{2}=(3,13),\text{
}\Sigma_{1}=0.4I_{2},\text{ }\Sigma_{2}=\left(
\begin{array}
[c]{cc}%
1.5 & -1.1\\
-1.1 & 1.5
\end{array}
\right)  ,
\]
In this case we take $\varepsilon=0.10$ and the outliers are generated with
uniform distribution in the square $[-50,5]\times\lbrack-50,5]$. The sample
size for this scenario is $n=1000$.  }

{\normalsize \noindent{\textbf{SideNoise2H}:} In this case we also consider
$K=2$ clusters with weights $\boldsymbol{\alpha}=(0.75,0.25)$ but this time
\ in $R^{20}$. The generating process for the first two coordinates is as in
the previous case, including the addition of outliers (only the first two
coordinates are contaminated). The remaining eighteen coordinates are
independent standard normal random variables. The sample size for this
scenario is $n=2000$.  }

{\normalsize \noindent{\textbf{SideNoise3}:} In this case we consider $K=3$
clusters, with weights $\boldsymbol{\alpha}$$=(0.28,0.33,0.39)$ \ in  $R^{2}$.
The kernel distribution has multivariate distribution }

{\normalsize with
\[
\boldsymbol{\mu}_{1}=(-2,-2),\quad\boldsymbol{\mu}_{2}=(7,1).\quad
\boldsymbol{\mu}_{3}=(15,19),\quad
\]%
\[
\Sigma_{1}=\left(
\begin{array}
[c]{cc}%
1 & 0.5\\
0.5 & 1
\end{array}
\right)  ,\quad\Sigma_{2}=\left(
\begin{array}
[c]{cc}%
2 & -1.5\\
-1.5 & 2
\end{array}
\right)  ,\quad\Sigma_{3}=\left(
\begin{array}
[c]{cc}%
2 & 1.3\\
1.3 & 2
\end{array}
\right)  .
\]
In this case the fraction of contamination is $\varepsilon=0.10$ with uniform
distribution in the rectangle $[-20,15]\times\lbrack-50,5]$. The sample size
for this scenario is $n=1000$. }

{\normalsize \noindent{\textbf{RandomScatter}:} In this case we have $K=6$
clusters with weights
\[
\boldsymbol{\alpha}=(1/11,2/11,2/11,2/11,2/11,2/11).
\]
in $R^{2}$. The kernel distribution is normal and $\boldsymbol{\mu}%
_{k}=3(k-3)(1,1),1\leq k\leq6.$ For each \ replication $\Sigma_{k}=U_{k}%
U_{k}^{\text{T}}$, where $U_{k}$ is a $2\times2$ random matrix, whose elements
are independent uniform random variables on $[-1,1]$. The fraction of
contamination is $\varepsilon=0.05$ generated from a uniform distribution on a
region obtained as follows. We first expand by a factor of two the smallest
box that contains the clean data. The sample size for this scenario is
$n=1200$.\  }

{\normalsize \noindent{\textbf{RandomScatterH}:} The observations are
generated as in the previous case but now with $p=10$, and $\boldsymbol{\mu
}_{k}=3(k-3)\mathbf{1},1\leq k\leq6$, where $\mathbf{1}$ is a vector of $10$
ones. Moreover the $U_{k}$s are of dimension $10\times10$. The sample size is
$n=1200$.
}

{\normalsize \noindent{\textbf{Clean Gaussian data}}: several clean Gaussian
data scenarios are defined by choosing $\varepsilon=0$ in the various settings
described above. }

{\normalsize \textbf{Note. }For all the contaminated scenarios, if some
generated outliers fall inside the $99\%$ probability ellipsoids of the
distributions used to generate the clean clusters, then these data points are
removed and replaced by new ones generated in the same way until they all fall
outside the $99\%$ probability ellipsoids. }

\subsection{{\protect\normalsize \noindent{\textbf{ }}Estimators Compared in
the Simulation Study}}

{\normalsize \noindent\textbf{RMBC:} This is the clustering procedure that we
propose based on the estimators described in \ Sections \ref{algorithm normal}
and \ref{Allocation to clusters}. This procedure is applied using the function
RMBC in the package \textbf{RMBC} with all the default parameters. This
package can be downloaded from
https://github.com/jdgonzalezwork/RMBC\smallskip. }

{\normalsize \noindent\textbf{otrimle:} This approach was proposed by
\cite{corettoHennig2016}. The outliers are identified by adding a cluster with an
improper uniform density with level parameter $\delta$, \ \ \ \
\[
g_{\delta}\left(  \mathbf{x},\boldsymbol{\theta}\right)  =\alpha_{0}%
\delta+\sum_{j=1}^{K}\alpha_{j}f\left(  \mathbf{x},\boldsymbol{\mu}_{j}%
,\Sigma_{j}\right) , \ \ \ \ \sum_{j=0}^{k}\alpha_{j}=1.
\]
The first term in the mixture, $\delta$, can be interpreted as an outlier
generating improper density. The estimator $\hat{\boldsymbol{\theta}}_{\delta
}$ maximizes the pseudo likelihood of the sample, that is,
\[
\hat{\boldsymbol{\theta}}_{\delta}=\arg\max_{\boldsymbol{\theta}}\prod
_{i=1}^{n}g_{\delta}\left(  \mathbf{x}_{i},\boldsymbol{\theta}\right)  .
\]
\ The pseudo maximum likelihood estimator is computed using an approach
similar to the EM algorithm. Once the estimators, $(\widehat{\alpha}%
_{j},\widehat{\boldsymbol{\mu}}_{j},\widehat{\Sigma}_{j}),1\leq j\leq K$ and
$\widehat{\delta}$ are computed, the probability that each observation belongs
to a given cluster is obtained as if we were dealing with all true densities.
Each observation $\mathbf{x}_{i\text{ }}$ is assigned to the cluster with
largest posterior probability. An observation is called an outlier if it is
assigned to the pseudo uniform distribution. The R package \texttt{otrimle}
(based on an algorithm proposed in \cite{CHjmlr}) estimates all the parameters
including $\delta$. \medskip}

{\normalsize \noindent\textbf{mclust:} \cite{fraley2002model} proposed the
maximum likelihood estimator for the Gaussian mixture model with the
maximization
carried out by the EM algorithm, introduced by \cite{dempster1977maximum}. In
our simulation and example we use the function Mclust in the R-package mclust
described in \cite{scrucca2016mclust}. This recent implementation allows for
several constraints on the covariance matrices, such as equal volume or
similar shape.
\medskip}

{\normalsize \noindent\textbf{tclust:} \cite{garcia2008general} proposed the
$\alpha$\emph{-trimmed maximum likelihood estimators}, $0<\alpha<1$, which
maximizes the function
\[
\prod_{j=1}^{K}\prod_{i\in C_{j}}\alpha_{j}f(\mathbf{x}_{i},\boldsymbol{\mu
}_{j}\Sigma_{j}),
\]
where $C_{1},...C_{K}$ are disjoint subsets of $\{1,...,n\}$ such that if
$C_{0}=\{1,...,n\}-\cup_{j=1}^{K}C_{j}$ then $\#C_{0}=\varepsilon n$. The main
idea is that $\alpha n$ data points are collected in $C_{0}$ and labeled as
potential outliers, while the remaining $\mathbf{x}_{i}$ $\in C_{j}$ with
$j>0$ are regular observations.
This idea was previously explored by \ \cite{gallegos2005robust}, under the
assumption that the $\alpha_{i},$ $1\leq j\leq K$ \ are equal and all the
covariance determinants $|\Sigma_{i}|\ 1\leq j\leq K$ are also equal.
\cite{garcia2008general} study this estimator under a more general constraint
$\Gamma\leq\delta,$ where $\Gamma=\lambda_{max}/\lambda_{min}$ and
$\lambda_{max}$ and $\lambda_{min}$ are the maximum and minimum eigenvalues of
all the matrices \ $\Sigma_{j},j=1\dots,K$. This procedure is implemented in
{  the function tclust } in the package \textbf{tclust} built by
\cite{M5-DATA}. In our simulation we consider two versions of this procedure:
tclust with $\alpha= 0.05$ which is the default value for $\alpha$ in the
function tclust, and tclustOracle, where $\alpha$ matches the contamination
fraction in the given scenario. }

{\normalsize \noindent\textbf{Note:} The function tclust assigns the trimmed
observations to a separate cluster called $G_{0}$. For a fair evaluation of
the cluster results we re-assign these observations to the clusters with
largest estimated probability (see equation (\ref{perte})) calculated using
the parameter estimates reported by the function tclust. Moreover, the
outliers are also flagged using the confidence $99.9\%$ ellipsoids described
at the end of Section 5, again using the reported parameter estimates. }

{\normalsize \bigskip}

{\normalsize All the considered procedures are run using the default values
for their tuning parameters, except for tclustOracle, where the trimming
parameter is set equal to the true contamination level in the given scenario.
}

\subsection{{\protect\normalsize Performance Measures}}

{\normalsize \label{performancemeasures} }

{\normalsize \textbf{Misclassification Rate (MCR):} This measure focuses on
the identification of the true clusters. Suppose we have $n$ observations
known to belong to $K$ clusters labeled $1,2,...K$. Suppose that we run a
clustering algorithm to estimate these $K$ clusters. We compare each estimated
cluster with each one of the true clusters and search for the matching that
produces the minimum number, $m$, of misclassified items. Then the MCR is defined as
$MCR=m/n$. }

{\normalsize \noindent\textbf{Kullback--Leibler divergency    	 (KLD): } This divergence is computed   between the estimated and
true mixture densities, taking the true densitity as reference. }

{\normalsize \noindent\textbf{Sensitivity: } This measure evaluates the
procedure ability to flag the true outliers.  An observation is flagged as  outlier if it
falls  outside of all the 0.999 confidence ellipsoides calculated using the
estimated location and scatter parameters. Then, the  sensitivity is  defined as
the proportion of actual outliers that are flagged as such.  }

\subsection{{\protect\normalsize Simulation results}}

{\normalsize For each procedure, scenario and replication we compute the
performance measures described above. Table \ref{results without} gives the   
MRC and KL-divergence average performances for the case of clean Gaussian
data. Table \ref{results with} gives the MRC, KL-divergence and Sensitivity
average performances for the case of contaminated Gaussian data. }

\noindent\textbf{Results for Clean Gaussian Data}: The simulation results for
clean Gaussian data reveal the superiority of RMBC, which outperforms the other
robust procedures, in terms of both, MCR and KL-divergency. In this case,
mclust and tclustOracle are  designed for dealing with clean Gaussian
data. Therefore, it is  surprising  that RMBC outperform these two   procedures
in three scenarios (SunSpot5, RandomScatter and RandomScatterH). This may  be due to
the fact that RMBC uses a very good initial estimator.

\noindent\textbf{Results for Contaminated Gaussian Data}: {\normalsize Notice
that in general, as expected, tclustOracle outperforms tclust, for all the
metrics, whenever the tclust  default trimming parameter ($\alpha= 0.05$) falls below
the fraction of contamination in the simulated data. On the other hand tclust does
relatively well when the percentage of outliers is below $5\%$.
Unfortunately, the true fraction of contamination in the data is seldom known
and difficult to estimate from the data. }

{\normalsize Regarding MCR, RMBC has the best performance in all the
considered scenarios except for RandomScatterH where tclust has the best
performance. Even in this case RMBC is a relatively close runner up. Regarding
KLD, RMBC overall has a good performance. It comes first under all the
considered scenarios except for SideNoise2H and SideNoise3 where otrimle and
tclustOracle perform slightly better. Regarding sensitivity, RMBC does in
general very well, detecting always at least $96\%$ of the true outliers. As
expected, mclust fails to identify the greatest majority of the true outliers.
Also notice that tclust has poor sensitivity (fails to detect a large fraction
of true outliers) whenever its default trimming parameter ($\alpha=0.05$)
falls below the percentage of contamination in the data. }

{\normalsize
}

{\normalsize { \begin{table}[ptb]
{\normalsize { \centering
\ {\small
\begin{tabular}
[c]{lcccccc}\hline
method & SunSpot5 & SideNoise2 & SideNoise2H & SideNoise3 & RandomScatter &
RandomScatterH\\\hline
&  &  & \textbf{{MCR \%}} &  &  & \\
RMBC & $2.15$ & $0.13$ & $0.13$ & $0.16$ & $0.83$ & $0.34$\\
otrimle & $15.54$ & $2.97$ & $0.07$ & $0.59$ & $34.80$ & $20.75$\\
tclust & $5.33$ & $0.48$ & $0.33$ & $0.36$ & $7.05$ & $1.06$\\
tclustoracle & $4.53$ & $0.09$ & $0.08$ & $0.09$ & $5.61$ & $0.39 $\\
mclust & $6.56$ & $0.00$ & $0.00$ & $0.00$ & $3.77$ & $7.22$\\\hline
&  &  & \textbf{{KLD}} &  &  & \\
RMBC & $0.04$ & $0.01$ & $0.12$ & $0.02$ & $0.04$ & $0.21$\\
otrimle & $0.41$ & $2.69$ & $0.12$ & $0.01$ & $10.49 $ & $9.57$\\
tclust & $0.08$ & $0.05$ & $0.17$ & $0.03$ & $0.66$ & $2.21$\\
tclustoracle & $0.03$ & $0.01$ & $0.12$ & $0.01$ & $0.49$ & $2.15$\\
mclust & $0.03$ & $0.00$ & $0.12$ & $0.01$ & $0.09$ & $0.91$\\\hline
\par &  &  &  &  &  &
\end{tabular}
} }  }\caption{ Simulation results for the six scenarios and the different
clustering procedures without outliers. MCR: Misclassification Rate, KLD:
Kullback-Leibler Divergence.}%
\label{results without}%
\end{table}} }

{\normalsize { \begin{table}[ptb]
{\normalsize { \centering
\ {\small
\begin{tabular}
[c]{lcccccc}\hline
method & SunSpot5 & SideNoise2 & SideNoise2H & SideNoise3 & RandomScatter &
RandomScatterH\\\hline
&  &  & \textbf{{MCR \%}} &  &  & \\
RMBC & $2.19$ & $0.05$ & $0.04$ & $0.08$ & $1.29$ & $0.62$\\
otrimle & $15.41$ & $2.63$ & $0.14$ & $1.05$ & $31.48$ & $21.28$\\
tclust & $5.35$ & $5.75$ & $9.43$ & $12.46$ & $7.49$ & $0.39$\\
tclustoracle & $7.92$ & $0.09$ & $0.13$ & $0.12$ & $7.49$ & $0.39$\\
mclust & $13.84$ & $0.07$ & $24.01$ & $0.53$ & $11.06$ & $10.77$\\\hline
&  &  & \textbf{{KLD}} &  &  & \\
RMBC & $0.04$ & $0.03$ & $0.20$ & $0.06$ & $0.07$ & $0.36$\\
otrimle & $0.43$ & $1.28$ & $0.14$ & $0.02$ & $9.30 $ & $9.75$\\
tclust & $0.07$ & $1.62$ & $1.74$ & $0.63$ & $0.55$ & $2.19$\\
tclustoracle & $0.06$ & $0.10$ & $0.19$ & $0.02$ & $0.55$ & $2.19$\\
mclust & $0.12$ & $1.12$ & $2.11$ & $1.06$ & $0.41$ & $1.51$\\\hline
&  &  & \textbf{{Sensitivity \%}} &  &  & \\
RMBC & $100.00$ & $99.61$ & $99.33$ & $96.72$ & $96.59$ & $100.00$\\
otrimle & $100.00$ & $99.69$ & $99.44$ & $98.56$ & $98.25$ & $100.00$\\
tclust & $100.00$ & $79.26$ & $73.40$ & $63.14$ & $96.33$ & $100.00$\\
tclustoracle & $84.08$ & $99.45$ & $99.12$ & $98.10$ & $96.33$ & $100.00$\\
mclust & $16.91$ & $4.49$ & $0.05$ & $9.24$ & $0.11$ & $0.01$\\\hline
\par &  &  &  &  &  &
\end{tabular}
} }  }\caption{ Simulation results for the six scenarios and the different
clustering procedures with outlier contamination. MCR: misclassification Rate,
KLD: Kullback-Leibler Divergence.}%
\label{results with}%
\end{table}} }

\section{{\protect\normalsize Application to Real Data \label{real}}}

{\normalsize
Phytoplankton, being a primary producer, plays a fundamental role in the
marine ecosystem. Furthermore, there are some phytoplankton species that can
be used as biological indicators of pollution in oceanic areas, and others
that produce massive algal blooms that affect activities carried out by man.
So estimating phytoplankton abundance is an important ecological problem.  }

{\normalsize The acoustic monitoring of phytoplankton is a potentially useful
technique for estimating the abundance of these organisms in real time.
Therefore, in the last decade ultrasound techniques have been developed to
obtain information about these organisms. See for example
\cite{BlancAcousticLeter}, \cite{bok2010ultrasound} and
\cite{blanc2017ultrasonic}.  }

{\normalsize In particular we will work with data from
\cite{cinquini2016advances}, obtained by taking laboratory measurements of
ultrasonic acoustic signals: a pulse is emitted by a transducer, this pulse
interacts with phytoplankton suspended in the water and produces an acoustic
dispersion (scattering), which is recorded by an electronic acquisition
device.  }

\subsection*{{\protect\normalsize Description of the dataset}}

{\normalsize A filtering process of the signal is performed in a first stage.
Portions of the signal belong to one of the two main cases:  }

\begin{itemize}
\item {\normalsize (a) Signals corresponding to the acoustic response of
phytoplankton.  }

\item {\normalsize (b) Signals corresponding to spurious dispersers, such as
bubbles or particles in suspension, whose intensity is greater than in case
(a).  }
\end{itemize}

{\normalsize To classify a signal in one of these two groups biologists create
a vector $(X_{1},X_{2})$ defined as follows:  }

{\normalsize
\[
X_{1}=\mbox{ratio of filtered to non-filtered signal power},
\]
\[
X_{2}=\mbox{filtered signal power expressed in dB}.
\]
The available data consists of 375 such measurements (see Figure
\ref{figuraoUT}). These data is particularly useful to compare robust
procedures because 20\% of these measurements are known to be outliers
produced by a communication failure between the electronic device (digital
oscilloscope) and the software for acquiring the acoustic signal. This failure
occurs once every 5 microseconds, which allows the scientists to identify the
outliers. The outliers appear as a separated group in the region $X_{1}<0.5$
and $X_{2}>20$ in Figure \ref{figuraoUT}.  }

{\normalsize \begin{figure}[ptb]
\begin{center}
{\normalsize \includegraphics[scale=0.6]{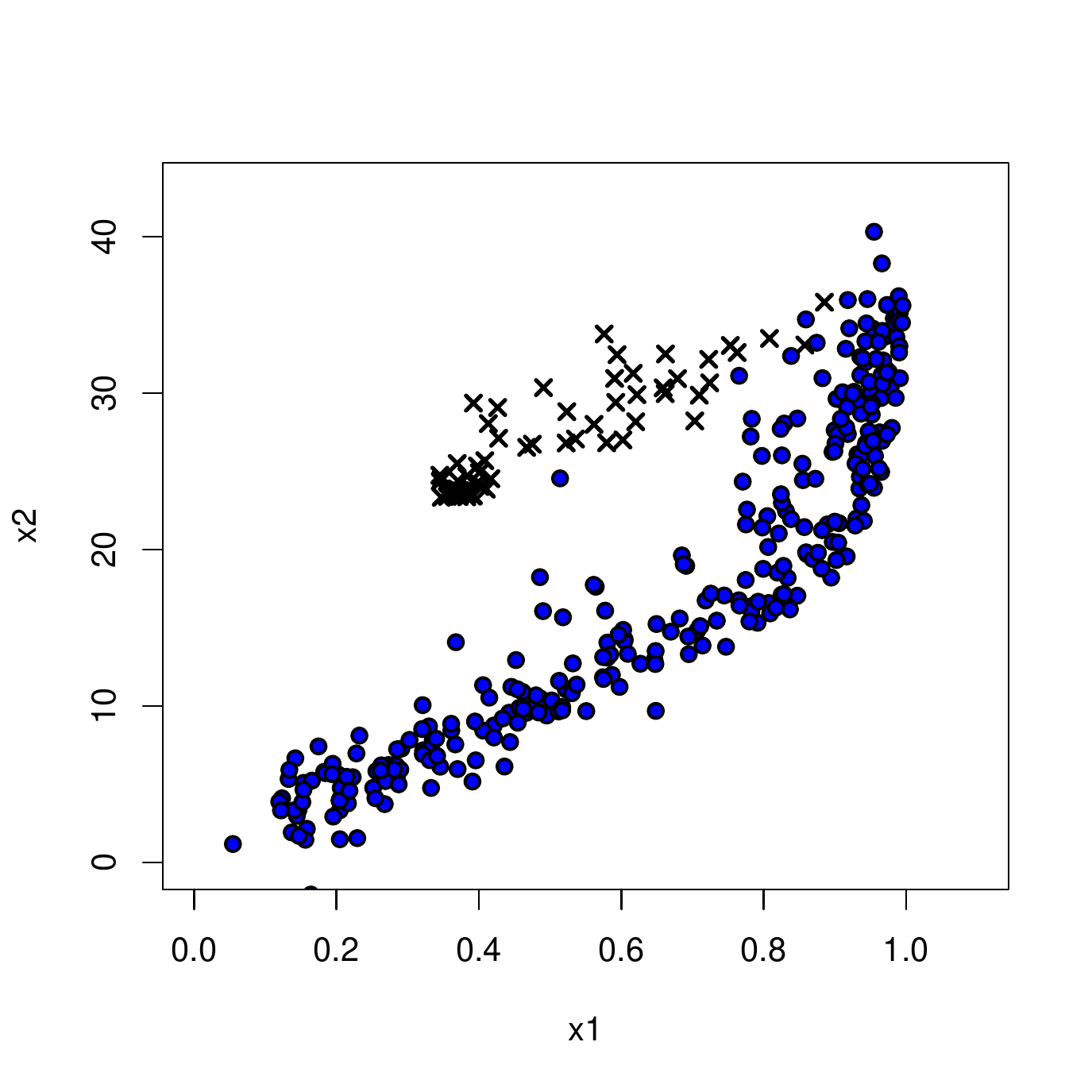}  }
\end{center}
{\normalsize  }\caption{Original Data ($n=375$). Circles and crosses correspond to regular observations and outliers respectively.}%
\label{figuraoUT}%
\end{figure} }

\subsection*{{\protect\normalsize Clustering analysis}}

{\normalsize
Now we apply the four model-based clustering procedures compared in our
simulation study to assess their ability to separate the observations of type
(a) and (b). The performance of the estimators is evaluated by the MCR, the Kullback-Leibler divergence  and the
sensitivity }

{\normalsize Since in this example the outliers are known, we can remove these
outliers and define the ``true groups'' as the partition produced by MCLUST
(the classical procedure) applied to the clean data. We call this ``the
reference partition''. Then, we apply the four clustering procedures to the
whole data set including the outliers. }

{\normalsize In the first panel of Figure \ref{DatosRealesFig} we show the
cleaned data obtained after the true outliers identified by the biologists are
removed. In this panel we also show the allocation of the observations to the
two clusters. By scientific prior knowledge we know that measurements of the
type (a) tend to have larger values of $X_{1}$ and $X_{2}$. Therefore, in the
partition of the clean measurements produced by MCLUST, we identify the data
points represented by triangles as measurements of type (a) and those
represented by circles as measurements of type (b). }

{\normalsize Table \ref{resultadosRealDataMixture} shows the performance
measures for the four considered procedures. Overall, RMBC has the best
performance. Otrimle and TCLUST with oracle tuning parameter $\ \ \ \ =0.2$
come second, except for KLD where they exhibit the worst performance. MCLUST,
TCLUST and Otrimle have zero sensitivity because they fail to flag the true
outliers. To reproduce Figure \ref{figuraoUT} and Table
\ref{resultadosRealDataMixture} of this example link to \newline
https://github.com/jdgonzalezwork/RMBC\_{Reproducibility}. }


{\normalsize \begin{table}[tbh]
\begin{center}
{\normalsize
\begin{tabular}
[r]{lrrrr}\hline
& RMBC & TCLUST & MCLUST & Otrimle\\\hline
Misclassification Rate \%  & 6.33 & 7.67 & 18.67 & 7.67\\
Kullback-Leibler Divergence & 0.29 & 2.69 & 0.57 & 2.15\\
Sensitivity \% & 74.67 & 0.00 & 0.00 & 0.00 \\\hline
\end{tabular}

}
\end{center}
\caption{ Performance of the compared model-based clustering procedures
applied to the phytoplankton data. The reported values for TCLUST correspond
to the choice $\alpha=0.20$, the actual fraction of outliers in the data
(ORACLE).}%
\label{resultadosRealDataMixture}%
\end{table}\color{black} }

{\normalsize
}

{\normalsize \begin{figure}[ptb]
{\normalsize \includegraphics[scale=0.5]{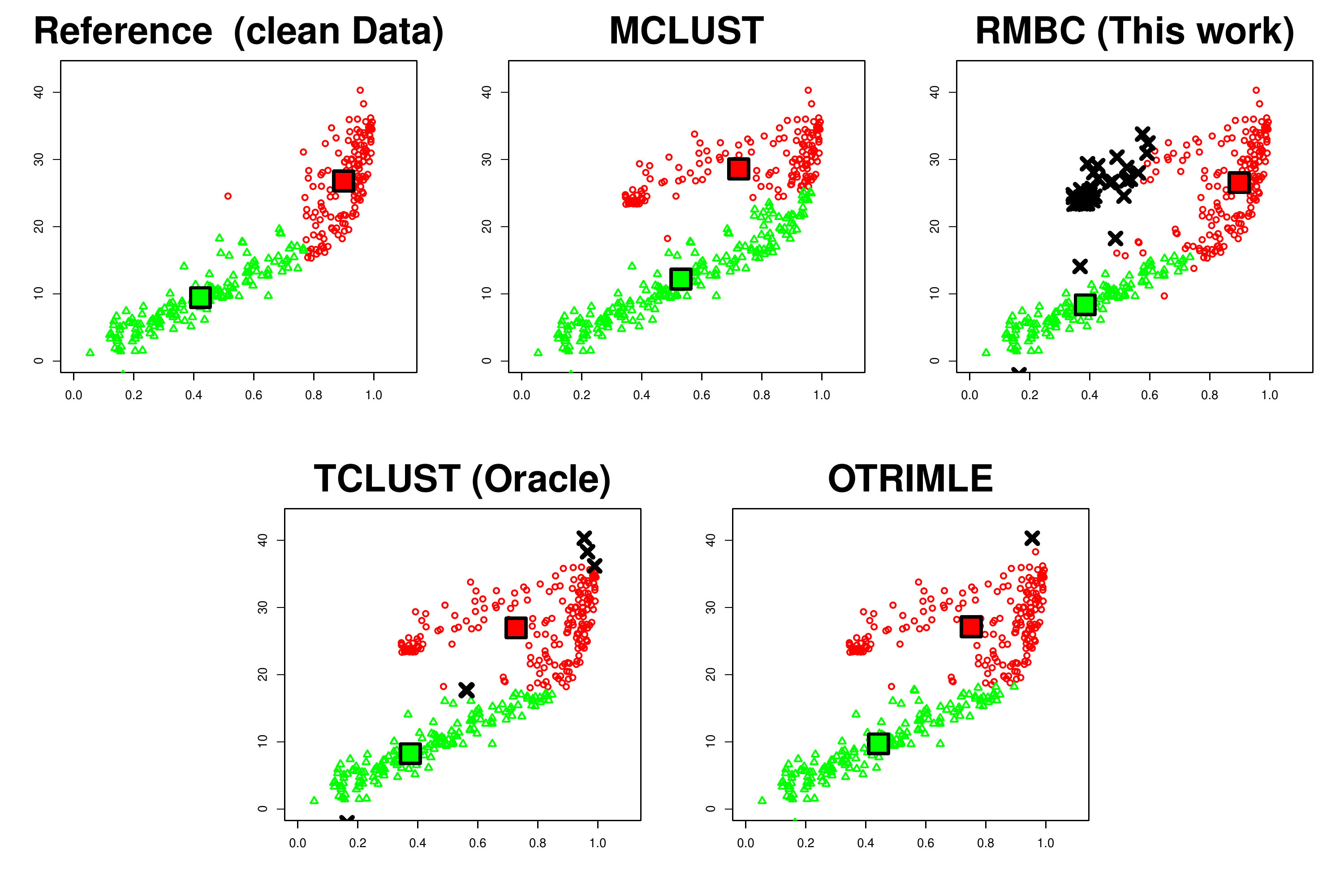}
}\caption{Results from the classic and robust model-based clustering procedures. The cluster of triangles and the cluster of circles correspond to observations of type (a) and (b) respectively. Crosses and squares represent detected outliers and cluster estimated centers respectively.}%
\label{DatosRealesFig}%
\end{figure}\newpage}

\section{{\protect\normalsize Conclusions}}

{\normalsize \label{conlusions} }

{\normalsize We present a general framework for the robust estimation of the
parameters of a mixture model and show how this can be used to perform robust
model-based clustering. Our proposal has some desirable features: }

\begin{itemize}
\item {\normalsize The procedure is Fisher consistent under mild regularity
assumptions. }

\item {\normalsize The procedure compares favorably with other robust and
nonrobust model-based clustering proposals in an extensive simulation study
and a real data application. }

\item {\normalsize The procedure can be applied using an efficient computing
algorithm implemented in the R-package \textbf{RMBC}.}

\item {\normalsize The procedure's tuning parameters do not depend on the
(usually unknown) fraction of outliers in the data. }
\end{itemize}

{\normalsize \smallskip}

\section*{{\protect\normalsize Acknowledgements}}

{\normalsize Juan D. Gonz\'{a}lez was partially supported by Grant
PIDDEF 02-2020, Program of the Argentinian Ministry of Defense, Victor J.
Yohai by grants \newline20020170100330BA from the Universidad de Buenos Aires
and PICT-201-0377 from ANPYCT, Argentina, and Ruben H. Zamar by the Natural
Sciences and Engineering Research Council of Canada. \bigskip}

\section*{{\protect\normalsize Appendix}}

{\normalsize {\large Proof of Theorem 1} }

{\normalsize
We must show that for $1\leq k\leq K,$
\begin{equation}
\mathcal{\ }\alpha_{0k}=E_{H_{0}}(\widetilde{\alpha}_{k}(\mathbf{x}%
,\boldsymbol{\alpha}_{0}\boldsymbol{,}\Theta_{0})) \label{prim}%
\end{equation}
and
\begin{equation}
\boldsymbol{\theta}_{0k}\ =\mathbf{g}\left(  \frac{E_{H_{0}}\left(
\widetilde{\alpha}_{k}(\mathbf{x},\boldsymbol{\alpha}_{0}\boldsymbol{,}%
\Theta_{0})\eta_{r}(\mathbf{x},\boldsymbol{\theta}_{0k})\right)  }{E_{H_{0}%
}(\widetilde{\alpha}_{k}(\mathbf{x},\boldsymbol{\alpha}_{0}\boldsymbol{,}%
\Theta_{0}))},...,\frac{E_{H_{0}}\left(  \widetilde{\alpha}_{k}(\mathbf{x}%
,\boldsymbol{\alpha}_{0}\boldsymbol{,}\Theta_{0})\eta_{h}(\mathbf{x}%
,\boldsymbol{\theta}_{0k})\right)  }{E_{H_{0}}(\widetilde{\alpha}%
_{k}(\mathbf{x},\boldsymbol{\alpha}_{0}\boldsymbol{,}\Theta_{0}))}\right)  .
\label{seec}%
\end{equation}
To prove (\ref{prim}) we write
\begin{align}
&  E_{H_{0}}(\widetilde{\alpha}_{k}(\mathbf{x},\boldsymbol{\alpha}%
_{0}\boldsymbol{,}\Theta_{0}))\nonumber\\
&  =\int_{-\infty}^{\infty}\cdots\int_{-\infty}^{\infty}\widetilde{\alpha}%
_{k}(\mathbf{x},\boldsymbol{\alpha}_{0}\boldsymbol{,}\Theta_{0})\sum_{l=1}%
^{K}\alpha_{0l}f_{\boldsymbol{\ }}(\mathbf{x},\boldsymbol{\theta}%
_{0l})d\mathbf{x}\nonumber\\
&  =\int_{-\infty}^{\infty}\cdots\int_{-\infty}^{\infty}\frac{f\left(
\mathbf{x},\boldsymbol{\theta}_{0k}\right)  \alpha_{0k}}{\sum_{l=1}^{K}%
\alpha_{0l}f_{\boldsymbol{\ }}(\mathbf{x},\boldsymbol{\theta}_{0l})}\sum
_{l=1}^{K}\alpha_{0l}f_{\boldsymbol{\ }}(\mathbf{x},\boldsymbol{\theta}%
_{0l})d\mathbf{x}\nonumber\\
&  =\int_{-\infty}^{\infty}\cdots\int_{-\infty}^{\infty}f\left(
\mathbf{x},\boldsymbol{\theta}_{0k}\right)  \alpha_{0k}d\mathbf{x=}\alpha
_{0k}. \label{eq22}%
\end{align}
To prove (\ref{seec}), by \ (\ref{eq20}) it is enough to show that fixing
$1\leq r\leq h$ and $1\leq k\leq K$ we have
\begin{equation}
\frac{E_{H_{0}}\left(  \widetilde{\alpha}_{k}(\mathbf{x},\boldsymbol{\alpha
}_{0}\boldsymbol{,}\Theta_{0})\eta_{r}(\mathbf{x},\boldsymbol{\theta}%
_{0k})\right)  }{E_{H_{0}}(\widetilde{\alpha}_{k}(\mathbf{x}%
,\boldsymbol{\alpha}_{0}\boldsymbol{,}\Theta_{0}))}=E_{F_{\boldsymbol{\theta
}_{0k}}}\left(  \eta_{r}(\mathbf{x},\boldsymbol{\theta}_{0k})\right)  .
\label{2333}%
\end{equation}
By (\ref{198}) and (\ref{eq22}) we get
\begin{align*}
\frac{E_{H_{0}}\left(  \widetilde{\alpha}_{k}(\mathbf{x},\boldsymbol{\alpha
}_{0}\boldsymbol{,\Theta}_{0})\eta_{r}(\mathbf{x},\boldsymbol{\theta}%
_{0k})\right)  }{E_{H_{0}}(\widetilde{\alpha}_{0k}(\mathbf{x}%
,\boldsymbol{\alpha}_{0}\boldsymbol{,\Theta}_{0}))}  &  =\frac{1}{\alpha_{0k}%
}\int_{-\infty}^{\infty}\cdots\int_{-\infty}^{\infty}\frac{f\left(
\mathbf{x},\boldsymbol{\theta}_{0k}\right)  \alpha_{0k}\eta_{r}(\mathbf{x}%
,\boldsymbol{\theta}_{0k})\sum_{l=1}^{K}\alpha_{0l}f_{\boldsymbol{\ }%
}(\mathbf{x},\boldsymbol{\theta}_{0l})}{\sum_{l=1}^{K}\alpha_{0l}%
f_{\boldsymbol{\ }}(\mathbf{x},\boldsymbol{\theta}_{0l})}d\mathbf{x}\\
&  =\frac{1}{\alpha_{0k}}\int_{-\infty}^{\infty}\cdots\int_{-\infty}^{\infty
}f\left(  \mathbf{x},\boldsymbol{\theta}_{0k}\right)  \alpha_{0k}\eta
_{r}(\mathbf{x},\boldsymbol{\theta}_{0k})d\mathbf{x}\\
&  =\int_{-\infty}^{\infty}\cdots\int_{-\infty}^{\infty}f\left(
\mathbf{x},\boldsymbol{\theta}_{0k}\right)  \eta_{r}(\mathbf{x}%
,\boldsymbol{\theta}_{0k})d\mathbf{x}\\
&  =E_{F_{\Theta_{0k}}}\left(  \eta_{r}(\mathbf{x},\boldsymbol{\theta}%
_{0k})\right)  ,
\end{align*}
proving (\ref{2333}).
}

{\normalsize To show that the S estimator functional fits the general
framework outlined in Section \ref{base0} we must show that this functional
satisfies a system of fixed point equations. To obtain the estimating
equations of the S functional we consider a minimization problem which is
equivalent to (\ref{defSestimador}) but free of side constraints. We introduce
the auxiliary functional $A$ defined as
\[
A(F,\boldsymbol{\mu},\Sigma)=|\Sigma|^{1/(2p)}\sigma(F,\boldsymbol{\mu}%
,\Sigma).
\]
The following lemmas establish the relationship between the functionals $S$
and $A$. }

\begin{lemma}
{\normalsize \label{multiplicacionporescalar} For all $\lambda>0$,
\begin{equation}
\sigma(F,\boldsymbol{\mu},\lambda\Sigma)=\sigma(F,\boldsymbol{\mu}%
,\Sigma)/\sqrt{\lambda} \label{homS}%
\end{equation}
and so
\begin{equation}
A(F,\boldsymbol{\mu},\lambda\Sigma)=A(F,\boldsymbol{\mu},\Sigma). \label{homA}%
\end{equation}
}
\end{lemma}

{\normalsize {\large Proof } }

{\normalsize
For any $\lambda>0$,
\begin{equation}
d(\mathbf{x},\boldsymbol{\mu},\lambda\Sigma)=\lambda^{-1/2}d(\mathbf{x}%
,\boldsymbol{\mu},\Sigma). \label{propMahana}%
\end{equation}
Note that $\sigma(F,$$\boldsymbol{\mu}$$,\lambda\Sigma)$ satisfies the
equation
\[
E_{F}\left(  \rho\left(  \frac{d(\mathbf{x},\boldsymbol{\mu},\lambda\Sigma
)}{\sigma(F,\boldsymbol{\mu},\lambda\Sigma)}\right)  \right)  =\frac{1}{2}.
\]
Applying (\ref{propMahana}) we get
\[
E_{F}\left(  \rho\left(  \frac{d(\mathbf{x},\boldsymbol{\mu},\Sigma)}%
{\sqrt{\lambda}\sigma(F,\boldsymbol{\mu},\lambda\Sigma)}\right)  \right)
=\frac{1}{2}.
\]
Then,
\begin{equation}
\sqrt{\lambda}\sigma(F,\boldsymbol{\mu},\lambda\Sigma)=\sigma
(F,\boldsymbol{\mu},\Sigma) \label{SS}%
\end{equation}
and (\ref{homS}) is proved. Now we we will show (\ref{homA})
\[%
\begin{array}
[c]{rl}%
A(F,\boldsymbol{\mu},\lambda\Sigma) & =|\lambda\Sigma|^{1/(2p)}\sigma
(F,\boldsymbol{\mu},\lambda\Sigma)\\
& =(\lambda^{p})^{1/(2p)}|\Sigma|^{1/(2p)}\frac{1}{\sqrt{\lambda}}%
\sigma(F,\boldsymbol{\mu},\Sigma)\\
& =|\Sigma|^{1/(2p)}\sigma(F,\boldsymbol{\mu},\Sigma)\\
& =A(F,\boldsymbol{\mu},\Sigma).
\end{array}
\]
}

\begin{lemma}
{\normalsize \label{Lemma2} Suppose that $(\boldsymbol{\mu}(F),\Sigma
(F))=\arg\min_{\sigma(F,\boldsymbol{\mu},\Sigma)=1}|\Sigma|$ and that $\ $%
\[
A(\boldsymbol{\mu}^{\ast}(F),\Sigma^{\ast}(F))=\min_{\Sigma}A(\boldsymbol{\mu
},\Sigma)
\]
}
\end{lemma}

{\normalsize Then $\boldsymbol{\mu}$$(F)=$$\boldsymbol{\mu}$$^{\ast}(F)$ and
$\ \Sigma(F)=\sigma(F,$$\boldsymbol{\mu}$$^{\ast}(F),\Sigma^{\ast}%
(F))^{2}\Sigma^{\ast}(F).$ }

{\normalsize {\large Proof } }

{\normalsize We shall show that he S functional $($$\boldsymbol{\mu}%
$$(F),\Sigma(F))$ also minimizes $A(F,$$\boldsymbol{\mu}$$,\Sigma)$ without
the constraint \ $\sigma(F,\boldsymbol{\mu},\Sigma)=1$. In fact, to minimize
$|\Sigma|$ is equivalent to minimizing $|\Sigma|^{1/2p}$. Therefore we have
\begin{equation}
\min_{\sigma(F,\boldsymbol{\mu},\Sigma)=1}|\Sigma|^{1/(2p)}=\min
_{\sigma(F,\boldsymbol{\mu},\Sigma)=1}|\Sigma|^{1/(2p)}\sigma
(F,\boldsymbol{\mu},\Sigma)=\min_{\sigma(F,\boldsymbol{\mu},\Sigma
)=1}A(F,\boldsymbol{\mu},\Sigma). \label{equivunouno}%
\end{equation}
Then $($$\boldsymbol{\mu}$$(F),\Sigma(F))$ is a minimum of $A(F,$
$\boldsymbol{\mu}$$,\Sigma)$ subject to $\sigma(F,$$\boldsymbol{\mu}$%
$,\Sigma)=1$. Let $(\boldsymbol{\mu}^{\ast}(F),\Sigma^{\ast}(F))=\arg
\min_{\boldsymbol{\mu}.\Sigma>0}A(F,\boldsymbol{\mu},\Sigma).$By Lemma
{\ref{multiplicacionporescalar} we have }$A(F,\boldsymbol{\mu}^{\ast}%
,\Sigma^{\ast})=A(F,$$\boldsymbol{\mu}$$^{\ast},\ \sigma^{2}(F,\boldsymbol{\mu
}^{\ast},\Sigma^{\ast})\Sigma^{\ast}).$ Since by (\ref{SS}) $\sigma
(F,\boldsymbol{\mu}^{\ast},\ \sigma^{2}(F,\boldsymbol{\mu}^{\ast},\Sigma
^{\ast})\Sigma^{\ast})=1,$we have
\[
\arg\min_{\boldsymbol{\mu}, \Sigma>0}A(F,\boldsymbol{\mu},\Sigma)=\arg
\min_{\boldsymbol{\mu},|\Sigma|=1}A(F,\boldsymbol{\mu},\Sigma),
\]
and this proves the Lemma.
}

{\normalsize {\large Proof of Theorem 2} }

{\normalsize By Lemma \ref{Lemma2} it is enough to show that the critical
points of $A(F,$$\boldsymbol{\mu}$$,\Sigma)$ satisfy
\begin{equation}
\boldsymbol{\mu}=\frac{E_{F}\left(  W\left(  \frac{d(\mathbf{x}%
,\boldsymbol{\mu},\Sigma)}{\sigma(F,\boldsymbol{\mu},\Sigma)}\right)
\mathbf{x}\right)  }{E_{F}\left(  W\left(  \frac{d(\mathbf{\ x}%
,\boldsymbol{\mu},\Sigma)}{\sigma(F,\boldsymbol{\mu},\Sigma)}\right)  \right)
} \label{locfix}%
\end{equation}%
\begin{equation}
\Sigma=c\Sigma^{\ast}, \label{cosfix}%
\end{equation}
where
\begin{equation}
\Sigma^{\ast}=E_{F}\left(  W\left(  \frac{d(\mathbf{x},\boldsymbol{\mu}%
,\Sigma)}{\sigma(F,\boldsymbol{\mu},\Sigma)}\right)  (\mathbf{x}%
-\boldsymbol{\mu})(\mathbf{x}-\boldsymbol{\mu})^{\text{T}}\right)  .
\label{cosfix2}%
\end{equation}
Then, by Lemma \ref{multiplicacionporescalar}, $c=\sigma(F,$ $\boldsymbol{\mu
}$$^{{}},\Sigma^{\ast})^{2}$. \noindent The critical points of $A(F,$%
$\boldsymbol{\mu}$$,\Sigma)$ satisfy the equations
\[
\frac{\partial\sigma(F,\boldsymbol{\mu},\Sigma)}{\partial\boldsymbol{\mu}%
}=0,\frac{\partial A(F,\boldsymbol{\mu},\Sigma)}{\partial\Sigma}=0.
\]
Note that
\[
\ \frac{\partial d^{2}(x,\boldsymbol{\mu},\Sigma)}{\partial\boldsymbol{\mu}%
}=-2\Sigma^{-1}\mathbf{(x-\boldsymbol{\mu}),}%
\]%
\[
\frac{\partial d(x,\boldsymbol{\mu},\Sigma)}{\partial\boldsymbol{\mu}}%
=\frac{-\Sigma^{-1}(x-\boldsymbol{\mu})}{d(x,\boldsymbol{\mu},\Sigma)},
\]%
\[
\frac{\partial}{\partial\boldsymbol{\mu}}\left(  \frac{d(x,\boldsymbol{\mu
},\Sigma)}{\sigma(F,\boldsymbol{\mu},\Sigma)}\right)  =\frac{\frac
{-\Sigma^{-1}(x-\boldsymbol{\mu})}{d(x,\boldsymbol{\mu},\Sigma)}%
\sigma(F,\boldsymbol{\mu},\Sigma)-d(x,\boldsymbol{\mu},\Sigma)\frac
{\partial\sigma(F,\boldsymbol{\mu},\Sigma)}{\partial\boldsymbol{\mu}}}%
{\sigma^{2}(F,\boldsymbol{\mu},\Sigma)}.
\]
Implicit differentiation of $\sigma(F,$$\boldsymbol{\mu}$$,\Sigma)$ with
respect to $\boldsymbol{\mu}$ gives
\[
E_{F}\left(  \psi\left(  \frac{d(\mathbf{x},\boldsymbol{\mu},\Sigma)}%
{\sigma(F,\boldsymbol{\mu},\Sigma)}\right)  \frac{\frac{-\Sigma^{-1}%
(x-\boldsymbol{\mu})}{d(x,\boldsymbol{\mu},\Sigma)}\sigma(F,\boldsymbol{\mu
},\Sigma)-d(x,\boldsymbol{\mu},\Sigma)\frac{\partial\sigma(F,\boldsymbol{\mu
},\Sigma)}{\partial\boldsymbol{\mu}}}{\sigma^{2}(F,\boldsymbol{\mu},\Sigma
)}\right)  =0,
\]
where $\psi=\rho^{\prime}.$ \ Putting $\partial\sigma(F,$$\boldsymbol{\mu}%
$$,\Sigma)/\partial$$\boldsymbol{\mu}$$=0$ and multiplying both sides by
$-\sigma(F,$$\boldsymbol{\mu}$$,\Sigma)^{2}\Sigma$ we get
\[
E_{F}\left(  \psi\left(  \frac{d(\mathbf{x},\boldsymbol{\mu},\Sigma)}%
{\sigma(F,\boldsymbol{\mu},\Sigma)}\right)  \frac{(\mathbf{x}-\boldsymbol{\mu
})}{d(x,\boldsymbol{\mu},\Sigma)}\sigma(F,\boldsymbol{\mu},\Sigma)\right)
=0,
\]
and%
\[
E_{F}\left(  \frac{\psi\left(  \frac{d(\mathbf{x},\boldsymbol{\mu},\Sigma
)}{\sigma(F,\boldsymbol{\mu},\Sigma)}\right)  }{\frac{d(\mathbf{x}%
,\boldsymbol{\mu},\Sigma)}{\sigma(F,\boldsymbol{\mu},\Sigma)}}%
\ \mathbf{(\mathbf{x}-\boldsymbol{\mu})}\right)  =0.
\]
Setting $W(t)=\psi(t)/t$ we get
\[
E_{F}\left(  W\left(  \frac{d(\mathbf{x},\boldsymbol{\mu},\Sigma)}%
{\sigma(F,\boldsymbol{\mu},\Sigma)}\right)  \mathbf{(x-\boldsymbol{\mu}%
)}\right)  =0,
\]
or equivalently
\[
\boldsymbol{\mu}=\frac{E_{F}\left(  W\left(  \frac{d(\mathbf{x}%
,\boldsymbol{\mu},\Sigma)}{\sigma(F,\boldsymbol{\mu},\Sigma)}\right)
\mathbf{x}\right)  }{E_{F}\left(  W\left(  \frac{d(\mathbf{x,\boldsymbol{\mu
},\Sigma)}}{\sigma(F,\boldsymbol{\mu},\Sigma)}\right)  \right)  }.
\]
We now differentiate $A(F,$$\boldsymbol{\mu}$$,\Sigma)$ with respect to
$\Sigma.$ \ We will use the following results%
\begin{equation}
\frac{\partial}{\partial\Sigma}|\Sigma|=|\Sigma|\Sigma^{-1} \label{result1}%
\end{equation}
and
\begin{equation}
\frac{\partial}{\partial\Sigma}\mathbf{a}^{\text{T}}\Sigma^{-1}\mathbf{b}%
=-\Sigma^{-1}\mathbf{ab}^{\text{T}}\Sigma^{-1}. \label{res2}%
\end{equation}
Then ,
\[
\frac{\partial d(\mathbf{x},\boldsymbol{\mu},\Sigma)}{\partial\Sigma}%
=\frac{-\Sigma^{-1}(\mathbf{x-\boldsymbol{\mu})(x-\boldsymbol{\mu})^{\text{T}%
}}\Sigma^{-1}}{2d(\mathbf{x},\boldsymbol{\mu},\Sigma)}.
\]
Differentiating $\sigma(F,$$\boldsymbol{\mu}$$,\Sigma)$ with respect to
$\Sigma$ \ we get
\begin{equation}
E_{F}\left(  \psi\left(  \frac{d(\mathbf{x},\boldsymbol{\mu},\Sigma)}%
{\sigma(F,\boldsymbol{\mu},\Sigma)}\right)  \frac{\frac{-\Sigma^{-1}%
(\mathbf{x-\boldsymbol{\mu})(x-\boldsymbol{\mu})^{\text{T}}}\Sigma^{-1}%
}{2d(\mathbf{x},\boldsymbol{\mu},\Sigma)}\sigma(F,\boldsymbol{\mu}%
,\Sigma)-d(\mathbf{x},\boldsymbol{\mu},\Sigma)\frac{\partial\sigma
(F,\boldsymbol{\mu},\Sigma)}{\partial\Sigma}}{\sigma^{2}(F,\boldsymbol{\mu
},\Sigma)}\right)  =0. \label{fin}%
\end{equation}
Besides differentiating $A(F,$$\boldsymbol{\mu}$$,\Sigma)$ with respecting to
$\Sigma$ \ we get
\[
\frac{\frac{\partial\sigma(F,\boldsymbol{\mu},\Sigma)}{\partial\Sigma}}%
{\sigma(F,\boldsymbol{\mu},\Sigma)}+\frac{|\Sigma|\Sigma^{-1}\ }{2p|\Sigma
|}=0,
\]
and therefore%
\begin{equation}
\frac{\partial\sigma(F,\boldsymbol{\mu},\Sigma)}{\partial\Sigma}=-\frac{1}%
{2p}\Sigma^{-1}\ \sigma(F,\boldsymbol{\mu},\Sigma). \label{fin2}%
\end{equation}
Therefore replacing $\partial\sigma(F,$$\boldsymbol{\mu}$$,\Sigma
)/\partial\Sigma$ in (\ref{fin}) we get
\[
E_{F}\left(  \psi\left(  \frac{d(\mathbf{x},\boldsymbol{\mu},\Sigma)}%
{\sigma(F,\boldsymbol{\mu},\Sigma)}\right)  \frac{\frac{-\Sigma^{-1}%
(\mathbf{x}-\boldsymbol{\mu})(\mathbf{x}-\boldsymbol{\mu})^{\text{T}}%
\Sigma^{-1}}{2d(\mathbf{x},\boldsymbol{\mu},\Sigma)}\sigma(F,\boldsymbol{\mu
},\Sigma)+d(\mathbf{x},\boldsymbol{\mu},\Sigma)\frac{1}{2p}\Sigma^{-1}%
\ \sigma(F,\boldsymbol{\mu},\Sigma)}{\sigma^{2}(F,\boldsymbol{\mu},\Sigma
)}\right)  =0
\]
and
\[
E_{F}\left(  \psi\left(  \frac{d(\mathbf{x},\boldsymbol{\mu},\Sigma)}%
{\sigma(F,\boldsymbol{\mu},\Sigma)}\right)  \left(  \frac{-\Sigma
^{-1}(\mathbf{x}-\boldsymbol{\mu})(\mathbf{x}-\boldsymbol{\mu})^{\text{T}%
}\Sigma^{-1}}{2d(\mathbf{x},\boldsymbol{\mu},\Sigma)\sigma(F,\boldsymbol{\mu
},\Sigma)}+\frac{1}{2p}\frac{d(\mathbf{x},\boldsymbol{\mu},\Sigma)}%
{\sigma(F,\boldsymbol{\mu},\Sigma)}\Sigma^{-1}\ \right)  \right)  =0.
\]
Multiplying by $\Sigma$ to the left and to the right we obtain
\[
E_{F}\left(  \psi\left(  \frac{d(\mathbf{x},\boldsymbol{\mu},\Sigma)}%
{\sigma(F,\boldsymbol{\mu},\Sigma)}\right)  \frac{(\mathbf{x}-\boldsymbol{\mu
})(\mathbf{x}-\boldsymbol{\mu})^{\text{T}}}{2d(\mathbf{x},\boldsymbol{\mu
},\Sigma)\sigma(F,\boldsymbol{\mu},\Sigma)}-\frac{1}{2p}\frac{d(\mathbf{x}%
,\boldsymbol{\mu},\Sigma)}{\sigma(F,\boldsymbol{\mu},\Sigma)}\Sigma\ \right)
=0,
\]%
\[
E_{F}\left(  \psi\left(  \frac{d(\mathbf{x},\boldsymbol{\mu},\Sigma)}%
{\sigma(F,\boldsymbol{\mu},\Sigma)}\right)  \frac{(\mathbf{x}-\boldsymbol{\mu
})(\mathbf{x}-\boldsymbol{\mu})^{\text{T}}}{2d(\mathbf{x},\boldsymbol{\mu
},\Sigma)\sigma(F,\boldsymbol{\mu},\Sigma)}\right)  =E_{F}\left(  \psi\left(
\frac{d(\mathbf{x},\boldsymbol{\mu},\Sigma)}{\sigma(F,\boldsymbol{\mu}%
,\Sigma)}\right)  \frac{1}{2p}\frac{d(\mathbf{x},\boldsymbol{\mu},\Sigma
)}{\sigma(F,\boldsymbol{\mu},\Sigma)}\right)  \Sigma
\]%
\[
\frac{2p}{2\sigma^{2}(F,\boldsymbol{\mu},\Sigma)}E_{F}\left(  \psi\left(
\frac{d(\mathbf{x},\boldsymbol{\mu},\Sigma)}{\sigma(F,\boldsymbol{\mu}%
,\Sigma)}\right)  \frac{(\mathbf{x}-\boldsymbol{\mu})(\mathbf{x}%
-\boldsymbol{\mu})^{\text{T}}}{\frac{d(\mathbf{x},\boldsymbol{\mu},\Sigma
)}{\sigma(F,\boldsymbol{\mu},\Sigma)}}\right)  =E_{F}\left(  \psi\left(
\frac{d(\mathbf{x},\boldsymbol{\mu},\Sigma)}{\sigma(F,\boldsymbol{\mu}%
,\Sigma)}\right)  \frac{d(\mathbf{x},\boldsymbol{\mu},\Sigma)}{\sigma
(F,\boldsymbol{\mu},\Sigma)}\right)  \Sigma.
\]
Setting
\[
c=c(F,\boldsymbol{\mu},\Sigma)=\frac{2p}{2\sigma^{2}(F,\boldsymbol{\mu}%
,\Sigma)E_{F}\left(  \psi\left(  \frac{d(\mathbf{x},\boldsymbol{\mu},\Sigma
)}{\sigma(F,\boldsymbol{\mu},\Sigma)}\right)  \frac{d(\mathbf{x}%
,\boldsymbol{\mu},\Sigma)}{\sigma(F,\boldsymbol{\mu},\Sigma)}\right)  }%
\]
we get
\[
\Sigma=cE_{F}\left(  W\left(  \frac{d(\mathbf{x},\boldsymbol{\mu},\Sigma
)}{\sigma(F,\boldsymbol{\mu},\Sigma)}\right)  (\mathbf{x}-\boldsymbol{\mu
})(\mathbf{x}-\boldsymbol{\mu})^{\text{T}}\right)  .
\]
proving (\ref{cosfix}) and (\ref{cosfix2}).
}

\end{document}